\documentclass[a4paper]{article}

\usepackage{amsmath,amssymb,amsfonts,amsthm} 
\usepackage{graphicx}

\usepackage{verbatim}

\usepackage[margin=1cm,font=small]{caption}

\usepackage{layout} 

\numberwithin{equation}{section}

\input{Macros.tex}

\newcommand{\PP}{\mathbf{P}} 
\newcommand{\QQ}{\mathbf{Q}}

\begin{document}

\title{Energy transport through rare collisions}
\author{
Fran\c cois Huveneers\footnote{
CEREMADE, 
Universit\' e de Paris-Dauphine,
Place du Maréchal De Lattre De Tassigny,
75775 PARIS CEDEX 16 - FRANCE.
E-mail: huveneers@ceremade.dauphine.fr.
Supported by the European Advanced Grant Macroscopic Laws and Dynamical Systems (MALADY) (ERC AdG 246953)}
}

\maketitle 

\abstract{
\noindent
We study a one-dimensional hamiltonian chain of masses perturbed by an energy conserving noise.
The dynamics is such that, according to its hamiltonian part, 
particles move freely in cells and interact with their neighbors through collisions, made possible by a small overlap of size $\epsilon >0$ between near cells. 
The noise only randomly flips the velocity of the particles. 
If $\epsilon \rightarrow 0$, and if time is rescaled by a factor $1 / \epsilon $, 
we show that energy evolves autonomously according to a stochastic equation, 
which hydrodynamic limit is known in some cases.
In particular, if only two different energies are present, the limiting process coincides with the simple symmetric exclusion process.
}

\vspace*{\stretch{1}}

\pagebreak

\section{Introduction}\label{section: introduction}

Fourier's law asserts that the flux of energy in a bulk of material is proportional to the gradient of temperature:
\begin{equation*}
J \; = \; - \kappa (T) \, \nabla T, 
\end{equation*} 
where $\kappa$ is the conductivity of the material. 
While this phenomenological law is widely verified in practice, 
its derivation from a microscopic hamiltonian dynamics still remains a very challenging question \cite{bon}.  

One actually knows that integrable hamiltonian systems usually violate Fourier's law, 
as it is the case for ordered harmonic chains \cite{rie}, 
for the Toda lattice with equal masses \cite{tod}\cite{ber}, 
or for a system of one-dimensional identical particles interacting only through collisions (see Remark \ref{rem: unpinned dynamics} in Subsection \ref{subsec: remarks}).
In these three examples, ballistic transport of energy is observed.
Moreover, the conductivity of one-dimensional chains of oscillators conserving the total momentum (no pinning) is generally expected to diverge \cite{lep}.

On an other hand, Fourier's law have been derived starting from some purely stochastic dynamics, such as exclusion processes.
In these models, particles have fixed position on a lattice (pinning) and exchange their energy with their neighbors according to the value of some random variable.  
The simple symmetric exclusion process (SSEP) is one of them, 
and constitutes an easy example where the heat equation can be recovered from a microscopic dynamics \cite{kip}.
To obtain a first hamiltonian derivation of Fourier's law, 
it seems thus desirable to start with hamiltonian systems which might look as close as possible to such stochastic models. 
In the recent years, many progress have been made in this direction.

Let us consider a periodic lattice of $N$ masses, each of them being confined in a region of the space by a pinning potential. 
We assume that the strength of the interaction between near atoms is controlled by a small parameter $\epsilon \ge 0$.
When $\epsilon = 0$, there is no interaction, and we suppose that the dynamics of each single atom has good mixing properties (hyperbolic dynamics). 
Then, when the interaction is turned on $(\epsilon >0)$ but is still very small, 
energy should flow between atoms at a much slower rate than 
the rate at which every atom reaches its own equilibrium for a given energy.  
One expects therefore the evolution of energy to be similar to the one of a stochastic process. 

Starting from there, it has been proposed to derive Fourier's law in two steps, as described by Gaspard and Gilbert \cite{gas}.
First, one tries to show that the energy of the atoms becomes an autonomous Markov process in the limit $\epsilon \rightarrow 0$,
if time has been rescaled by a factor $\epsilon^{-\alpha}$
for some $\alpha>0$, in order to obtain a non-trivial limit (but no space rescaling). 
The process obtained after this first step corresponds to a mesoscopic description of the material. 
Second, one hopes that the hydrodynamic limit of the mesoscopic process can be shown to coincide with the heat equation. 

The first part of this program has by now been carried out rigorously
when the interaction is of the form $\epsilon V$, where $V$ is a smooth potential coupling nearest neighbor atoms. 
Liverani and Olla \cite{liv} first considered the case where the uncoupled dynamics is made of a hamiltonian part
perturbed by a stochastic noise, modeling an hyperbolic dynamics. 
Dolgopyat and Liverani \cite{dol} were then able to obtain a similar result starting from a purely hamiltonian dynamics. 
One has however not yet been able to identify rigorously the hydrodynamic limit of the obtained mesoscopic processes. 

The aim of this paper is to investigate the case where the interaction between atoms takes place through elastic collision rather than a smooth potential. 
In this context, the parameter $\epsilon$ represents the size of a small interaction zone where near particles can collide, 
and the limit $\epsilon \rightarrow 0$ corresponds thus to a limit of rare interactions, rather than weak interactions. 
The model we consider is one-dimensional, and the lack of hyperbolicity of the uncoupled dynamics is supplied by a stochastic noise (see Subsection \ref{subsec: the model}). 

The limiting process we obtain is described by (\ref{limit evolution measures1}-\ref{limit evolution measures2}) below, 
and belongs to a class to which presents probabilistic tools can be applied in order to derive the heat equation in the hydrodynamic regime, 
at least when the number of energies reachable by the system is finite \cite{kip}.
In the special case where only two different energies are present, we even recognize the generator of the SSEP.
The mesoscopic description of the dynamics is thus so simple in this model, that it leaves us some hope to go
beyond the slightly artificial two steps procedure followed here. 

The rest of this paper is organized as follows. 
In Section \ref{section: model and result}, we describe the dynamics at a non technical level, we state our result, and give some notations to be used throughout the text.
A detailed definition of the dynamics together with some elementary observations is to be found in Section \ref{section: def of the dynamics}. 
Section \ref{Section: proof of the theorem} contains the proof of Theorem \ref{theorem: principal result}, 
admitting Lemma \ref{lemma: fundamental technical result}, which proof is deferred to Section \ref{Section: proof of the main lemma}. 
The proof of this lemma constitutes actually the core of the argument. 
Section \ref{section: uncoupled dynamics} deals with the convergence to equilibrium for the uncoupled dynamics.
Finally, the two figures of the text are collected in Section \ref{section: figures}.

\section{Model and result}\label{section: model and result}

\subsection{The model}\label{subsec: the model}
\begin{figure}[h]
\begin{center}
\includegraphics[width=0.8\textwidth]{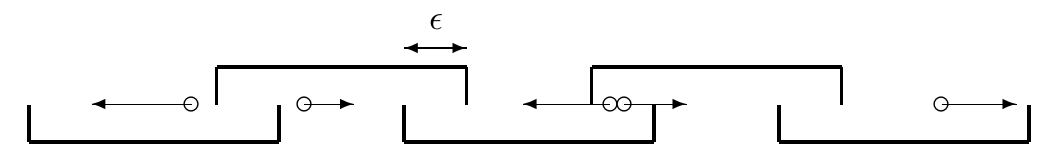}
\end{center}

\vspace{-0.8cm}

\caption{\label{figure: description physique de la chaine}
The dynamical system for $N=5$.
Particles may collide elastically with their neighbor when they enter in the overlap region of size $\epsilon$.
}
\end{figure}

We consider a system of $N$ classical point particles of unit mass, 
each one evolving in a one-dimensional cell of size one (see figure \ref{figure: description physique de la chaine} in Section \ref{section: figures}). 
Let us first describe the dynamics when all the particles move independently from each others. 
For this, let us associate a Poisson process (a "Poisson clock") to each of them. 
The $N$ Poisson processes are independent but have the same parameter $\lambda >0$. 
A particle moves then freely in its cell, meaning that it travels at a constant speed and is elastically reflected at the boundaries. 
In addition, each time its Poisson clock rings, the sign of its velocity is reversed. 
So the Poisson processes serve to model a chaotic dynamics inside the cells. 
Observe that the kinetic energy of every particle is a conserved quantity under this uncoupled dynamics. 

We then introduce some interaction between the particles. 
Let $0 < \epsilon < 1/2$, and put the cells in a one-dimensional row, as depicted in figure \ref{figure: description physique de la chaine} in Section \ref{section: figures}, 
so that there is an overlap of length $\epsilon$ between near cells.
If two nearest neighbor particles hit each others, they undergo an elastic collision, resulting in a exchange of their momenta. 
This is the only interaction ; 
in particular, particles do not hit against the boundary of neighbor cells.  
This dynamics still preserves the total kinetic energy of the system. 

To fully specify the dynamics, one finally needs to give the initial positions and velocities 
$x = \big(q_1, \dots , q_N,p_1,\dots ,p_N\big)$.
One supposes the initial positions to be such that particle $k$ is at the left of particle $k+1$ for $1 \le k \le N-1$.
The dynamics is so that this condition holds then at all further times.
If $q_k \in [0,1]$ for $1 \le k \le N$, this means $q_{k+1} \ge q_k -1 + \epsilon$ for $1 \le k \le N-1$.
One assumes also that no initial velocity is zero. 
For $t\ge 0$, the state of the system is then entirely given by the value of the Markov process
\begin{equation*}
X^\epsilon (x,t) = (q_1^\epsilon (x,t), \dots , q_N^\epsilon(x,t), p_1^\epsilon (x,t), \dots , p_N^\epsilon (x,t)),
\end{equation*}
where we have emphasized the dependence on $\epsilon$.

\subsection{The result}\label{subsec: the result}
Let $0 < \mathsf e_1 < \dots < \mathsf e_{N'} < + \infty$ be the initial energies, where $N'$ is an integer smaller or equal to $N$.
These are thus numbers such that 
$\frac{1}{2}p_k^2 \in \{ \mathsf e_1 , \dots , \mathsf e_{N'}\}$ for $1 \le k \le N$.
The dynamics is such that $\{\mathsf e_1, \dots , \mathsf e_{N'}\}$ constitute also the set of all energies reachable by the system for all times. 
One wants to describe how energy flows between the particles in the limit $\epsilon \rightarrow 0$. 
Since one suspects the rate of collisions to be proportional to $\epsilon$ (see Remark \ref{rem: scaling factor} in Subsection \ref{subsec: remarks}), 
one needs also to rescale time by a factor $\epsilon^{-1}$ in order to obtain a non-trivial limit. 
More precisely, we seek for the limit distribution of the time rescaled kinetic energy
\begin{equation*}
\mathcal E^\epsilon (t) 
=
\big( \mathcal E^\epsilon_1 (t) ,\dots ,  \mathcal E^\epsilon_N (t) \big)
=
\Big( \frac{1}{2} (p_1^\epsilon)^2(\epsilon^{-1}t), \dots , \frac{1}{2}(p_N^\epsilon)^2(\epsilon^{-1}t) \Big)
\end{equation*}
as $\epsilon \rightarrow 0$.

We need a couple of extra definitions to express our result:
\begin{enumerate}
\item
Let $\mathcal D ([0,1],\mathrm A)$ be the set of cad-lag functions on $[0,1]$ with value in $\mathrm A \subset\R^d$, 
and observe that actually $\mathcal E^\epsilon \in \mathcal D \big([0,1], \{ \mathsf e_1,\dots,\mathsf e_{N'}\}^N\big)$.

\item
Let $\gamma$ be a function on $\R_+^2$ given by
\begin{equation}\label{def of gamma}
\gamma (a,b)
=
\frac{1}{2}\max \{ \sqrt{2 a}, \sqrt{2b}\}.
\end{equation}

\item
Let $\mathbf Q^\epsilon$ be the physical space accessible to the particles: 
\begin{equation}\label{def of Q epsilon}
\mathbf Q^\epsilon \; = \; \{ (q_1,\dots ,q_N)\in\ [0,1]^N \; : \; q_{k+1} \,\ge\, q_k - 1 + \epsilon \; \; \text{for} \;\; 1 \le k \le N-1 \}.
\end{equation}

\item
A number $\xi\in\R$ is called diophantine if there exist constants $\mathrm C,\beta <+\infty$ such that, for every $p/q\in\Q$, one has
\begin{equation}
\Big| \xi - \frac{p}{q} \Big|
\; \ge \; \frac{\mathrm C}{q^\beta}.
\end{equation}
\end{enumerate}
With the definitions and notations introduced up to now, one has
\begin{Theorem}\label{theorem: principal result}
Let $e_{1}(0), \dots , e_{N}(0)\in \{ \mathsf e_1, \dots , \mathsf e_{N'}\}$.
Let us assume that
\begin{enumerate}
\item for $1 \le j \ne k \le N' $, the numbers $\sqrt{\mathsf e_j/\mathsf e_k}$ are diophantine,
\item the law of $X^\epsilon (\argument, 0)$ is the only probability measure which density is proportional to
\begin{equation*}
\chi_{\mathbf Q^{\epsilon}} (q_1, \dots , q_N ) \multiplication \chi_{( e_1(0), \dots ,e_N(0))} (p_1^2/2,\dots ,p_N^2/2).
\end{equation*}
\end{enumerate}
Then
\begin{enumerate}
\item as $\epsilon \rightarrow 0$, the distribution of the process $\mathcal E^\epsilon$ tends weakly to the distribution of a process
\[\mathcal E=(\mathcal E_1,\dots,\mathcal E_N)\; \in \; \mathcal D \big([0,1], \{ \mathsf e_1,\dots ,\mathsf e_{N'}\}^N\big),\]
\item  $\mathcal E$ is the unique Markov process which law $\overline\Proba $ solves the equations
\begin{align}
\partial_t \overline\Proba (e_1, \dots ,e_N, t)
=&\;
\sum_{k=1}^{N-1}\gamma (e_k,e_{k+1})
\Big( 
\overline\Proba (\dots , e_{k+1},e_k,\dots, t) - \overline\Proba (\dots , e_k, e_{k+1},\dots ,  t)
\Big), \label{limit evolution measures1}\\
\overline\Proba (e_1,\dots , e_N,0)
=&\;
\chi_{( e_1(0), \dots e_N(0) )}(e_1, \dots , e_N). \label{limit evolution measures2}
\end{align} 
\end{enumerate}
\end{Theorem}

\subsection{Remarks}\label{subsec: remarks}

\Remark\label{rem: unpinned dynamics}
The dynamics, with or without noise, should actually becomes completely elementary if the particles were not confined into cells\footnote{
we thank J.L. Lebowitz for having recalled us this fact}. 
Indeed, in this case, the system can be seen as a collection of $N$ independent particles, 
since they just exchange their momentum when colliding. 

\Remark\label{rem: initial conditions}
The condition on the initial measure is much more restrictive than needed. 
A close look at the proof shows that, with some extra work, one should be able to start from a measure with bounded density with respect to the positions.
Our choice mainly allows us to avoid an extra initial step that should be needed to apply Lemma \ref{lemma: fundamental technical result} below.

\Remark\label{rem: scaling factor}
It is not so obvious that time has to be rescaled by a factor $\epsilon^{-1}$ in order to obtain a non-trivial result 
(the correct scaling factor is $\epsilon^{-2}$ in \cite{dol} and \cite{liv}).
So let us try to motivate this a little bit, following the heuristic of \cite{gas}. 
It is enough to concentrate on the case of only two particles $(N=2)$, with two energies $\mathsf e_1 < \mathsf e_2$. 
When $\epsilon$ becomes very small, the rate of collisions between these particles should tend to zero as well. 
So, since the dynamics of uncoupled particles have good mixing properties, 
the distribution of position and velocity of a given particle with energy 
$e\in \{\mathsf e_1,\mathsf e_2\}$ should be well approximated by the equilibrium measure of an uncoupled particle of the same energy. 
It is easily checked that this measure is given by 
\begin{equation*}
\mu_e (u)
\; =\;
\frac{1}{2} \sum_{p=\pm\sqrt{2e}} \int_0^1 u (q,p) \, \dd q.
\end{equation*} 
Assuming that, at a given time, the energy of particle 1 is $\mathsf e_1$, and the energy of particle 2 is $\mathsf e_2$, 
it seems then reasonable to identify the instantaneous energy exchange rate as 
\begin{align*}
\tilde\gamma
\; = 
&\;
\lim_{\eta\rightarrow 0}\frac{1}{\eta}\,\Proba \big( \,\text{particle 1 hits particle 2 in a time interval }\eta\, \big) \\
\simeq&
\;
\lim_{\eta \rightarrow 0}\frac{1}{\eta}\;
\frac{1}{4} \hspace{-0.2cm} \sum_{\substack{p_1 = \pm \sqrt{2\mathsf e_1} \\ p_2 = \pm \sqrt{2\mathsf e_2} }}
\int_0^1 \dd q_1 \int_{1-\epsilon}^{2 - \epsilon} \dd q_2 \, 
\chi_{[0,(p_1-p_2)\eta]}(q_2-q_1) \, \chi_{\R_+}(p_1 - p_2), 
\end{align*}
where one has taken into account the fact that particle 1 should be at the left of particle 2. 
One computes that 
\begin{equation*}
\tilde\gamma
\; =\;
\frac{\epsilon \, \sqrt{2 \mathsf e_2}}{2}
\; = \;
\frac{\epsilon}{2} \max \{ \sqrt{2 \mathsf e_1}, \sqrt{2 \mathsf e_2} \}, 
\end{equation*}
so that indeed $\tilde\gamma$ will coincide with $\gamma$ as defined in \eqref{def of gamma} if time is rescaled by a factor $\epsilon^{-1}$.

\Remark\label{rem: diophantine condition}
The presence of a diophantine condition on the velocities may appear as a surprise in the present context. 
To see where it comes from, let us again take the case $N=N' = 2$, 
and let us assume that one velocity has value 1, and the other 2, violating thus the diophantine condition.
The dynamics of a single particle is such that, for any fixed time interval, in the microscopic time scale, there is a strictly positive probability, independent of $\epsilon$, 
that the Poisson clock of the particle does not ring during this time, so that its trajectory is in fact deterministic.      
Now, let us suppose that the two particles collide, 
and let us take the "typical" case where the collision does not occur too close to the borders of the overlap region between the two near cells. 
It can then be seen that,  no matter how small $\epsilon$ is taken, they will collide once more with each other, in a time interval of order 1, 
if they travel along the deterministic trajectory. 
This happens thus with a strictly positive probability. 

However, the basic assumption behind our theorem is that 
successive shocks occur "randomly", and are typically spaced by time intervals of order $\epsilon^{-1}$, in the microscopic time scale. 
This needs to be so for the coefficient $\gamma$ to be given by \eqref{def of gamma} in \eqref{limit evolution measures1}.
But we see that, if the quotients of velocities are rational like here, successive shocks are actually quite correlated, and tend to occur in clusters. 
It seems clear however, that these clusters should themselves occur at random and be typically spaced by time intervals of order  $\epsilon^{-1}$,
so that we conjecture Theorem \ref{theorem: principal result} to still remain valid without diophantine condition,
but with a smaller coeffeicient $\gamma$ since recollisions tend to cancel the transfer of energy.
We have not been able to show this.

\subsection{Notations}\label{subsection: notations}

\noindent\textbf{Constants.}
The number called constants in this text may depend on the following parameters of our problem : 
the number $N$ of particles, the number $N'$ of different energies, and the value of the smallest and the largest energy, namely $\mathsf e_1$ and $\mathsf e_{N'}$.  
They never depend nor on $\epsilon$, nor on time. 

\noindent\textbf{Time discretization ($\eta$).}
We will find convenient to divide time in small intervals. 
Throughout the text, we will denote these time steps by $\eta$. 
One always assume $\eta \le \epsilon$. 
Its exact value matters few, and one can always think that $\eta << \epsilon$.    

\noindent\textbf{Deterministic dynamics ($\sim$).}
Objects with a tilde refer to the deterministic dynamics. So e.g. $\tilde X^\epsilon$ is the trajectory of our process driven by the deterministic dynamics. 

\noindent\textbf{The phase space.}
Points of the phase space $\Phi$ are written $x = (\mathbf q, \mathbf p) = (q_1, \dots , q_N,p_1,\dots ,p_N)$. 
Moreover, one writes $\mathbf e = (e_1, \dots, e_n) = (p_1^2 /2, \dots , p_N^2/2)$.

\noindent\textbf{Norms.}
Let $\mathsf G$ be the Gibbs measure defined in Subsection \ref{subsec: Invariant measure} below.
The norm of the space $\Lp^p (\Phi,\mathsf G)$ is written $\| \argument \|_p$ ($1 \le p \le + \infty$).
If $\mu$ is a measure on $\Phi$, one writes $\| \mu \|_1 = \sup_{u: \| u \|_\infty \le 1} |\mu (u)|$.

\noindent\textbf{The space $\mathcal S^{\epsilon}$.}
Functions $u$ on $\Phi$ of the form $u (\mathbf q, \mathbf p) = \chi_{\mathbf Q^\epsilon} (\mathbf q) \multiplication v (\mathbf e)$ 
are said to belong to the space $\mathcal S^\epsilon$. 
A measure $\mu$ on $\Phi$ which density w.r.t. $\mathsf G$ belongs to $\mathcal S^\epsilon$, is itself said to belong to $\mathcal S^\epsilon$.

\section{The dynamics}\label{section: def of the dynamics}

\subsection{Definition}\label{subsection: def of the dynamics}

Let $\epsilon \ge 0$, let $N \ge N' \ge 1$ be two integers, and let  $0 < \mathsf e_1 < \dots < \mathsf e_{N'} < + \infty$ be given energies. 
Let 
\begin{equation*}
\PP 
\; = \;
\{ -\sqrt{2 \mathsf e_1}, \sqrt{2 \mathsf e_1} , \dots ,  -\sqrt{2 \mathsf e_{N'}}, \sqrt{2 \mathsf e_{N'}} \} 
\end{equation*}
be the set of velocities, let 
\begin{equation*}
\Phi
\; = \; 
[0,1]^N \times \PP^N
\end{equation*}
be the phase space,
and let 
\begin{equation*}
\Phi^\epsilon \; = \;  \mathbf Q^\epsilon \times \PP^N, 
\end{equation*}
with $\mathbf Q^\epsilon$ defined by \eqref{def of Q epsilon}, be the subset of $\Phi$ where the dynamics actually takes place.
Let $x = (\mathbf q, \mathbf p) = (q_1, \dots ,q_N, p_1,\dots ,p_N)\in \Phi$
and, for $t\ge 0$, let 
\begin{equation*}
X^\epsilon(x,t) 
=
(\mathbf q^\epsilon (x,t),\mathbf p^\epsilon (x,t))
=
( q^\epsilon_1 (x,t), \dots ,  q^\epsilon_N (x,t), p^\epsilon_1 (x,t), \dots , p^\epsilon_N (x,t))
\end{equation*}
be the stochastic process on $\Phi$ such that $X^\epsilon(x,0)=x$ and that, dropping out the superscript $\epsilon$ for clarity,
\begin{align}
\dd q_k 
\; = &\;
p_k \, \dd t, \label{equation of the motion 1} \\
\dd p_k
\; = &\;
-2p_k (\dd \mathrm N_k + \dd \mathrm N_k^{\mathrm l} + \dd \mathrm N_k^{\mathrm r})
+ (p_{k-1} - p_k)\, \dd \mathrm N^{\mathrm c}_{k-1} 
- (p_{k} - p_{k+1})\, \dd \mathrm N^{\mathrm c}_{k} \label{equation of the motion 2}
\end{align}
for $1 \le k \le N$, where
\begin{enumerate}
\item
the processes $\mathrm N_k$ are independent Poisson processes with identical parameter $\lambda >0$,

\item
the processes $\mathrm N_k^{\mathrm l}$ are cad-lag processes such that $\mathrm N_k^{\mathrm l} (t_+) - \mathrm N_k^{\mathrm l} (t_-) \in \{ 0,1\}$ for $t \ge 0$, 
with
\begin{equation*}
\mathrm N_k^{\mathrm l} (t_+) - \mathrm N_k^{\mathrm l} (t_-) \; = \; 1 
\quad \text{if and only if} \quad
q_k(t_-) = 0 \;\; \text{and}\;\; p_k(t_-) < 0,
\end{equation*}

\item
the processes $\mathrm N_k^{\mathrm r}$ are cad-lag processes such that $\mathrm N_k^{\mathrm r} (t_+) - \mathrm N_k^{\mathrm r} (t_-) \in \{ 0,1\}$ for $t \ge 0$, 
with
\begin{equation*}
\mathrm N_k^{\mathrm r} (t_+) - \mathrm N_k^{\mathrm r} (t_-) \; = \; 1 
\quad \text{if and only if} \quad
q_k(t_-) = 1 \;\; \text{and}\;\; p_k(t_-) > 0,
\end{equation*}

\item
the processes $\mathrm N_k^{\mathrm c}$ are cad-lag processes such that $\mathrm N_k^{\mathrm c} (t_+) - \mathrm N_k^{\mathrm c} (t_-) \in \{ 0,1\}$ for $t \ge 0$, 
with
\begin{equation*}
\mathrm N_k^{\mathrm c} (t_+) - \mathrm N_k^{\mathrm c} (t_-) \; = \; 1 
\quad \text{if and only if} \quad
q_{k+1}(t_-) = q_k (t_-) - 1 + \epsilon  \;\; \text{and}\;\; p_{k+1}(t_-) < p_k(t_-)
\end{equation*}
for $1 \le k \le N-1$, and $\mathrm N_N^{\mathrm c} = 0$.
\end{enumerate}
Here, $f (t_+)$ and $f (t_-)$ represent respectively the left and right limit of a cad-lag function $f$ at time $t$.

\begin{figure}[h]
\begin{center}
\includegraphics[width=0.48\textwidth]{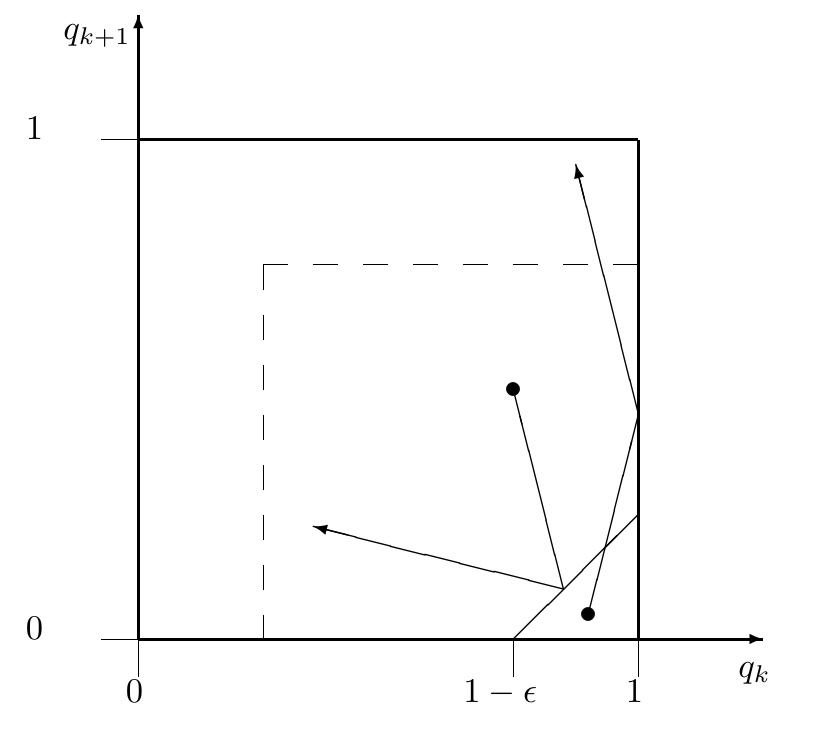}
\end{center}

\vspace{-0.8cm}

\caption{\label{figure: projection sur un plan a deux variables}
Projection of $\Phi$ on the $(q_k,q_{k+1})-$plane: 
the two circles represent projections of two possible states of the $N$-particle system.  
Points in the triangle in the lower right corner belong to $\Phi - \Phi^{\epsilon}$, since $q_{k+1} < q_{k} - 1 + \epsilon$ there. 
In the region on the left of the dotted line, particle $k$ may collide with particle $k-1$, 
whereas in the region above the dotted line, particle $k+1$ may collide with particle $k+2$.
}
\end{figure}

Equations (\ref{equation of the motion 1}-\ref{equation of the motion 2}) describe the evolution of a particle in a $N$-dimensional billiard, 
which projection on the $(q_k,q_{k+1})-$plane is depicted on figure \ref{figure: projection sur un plan a deux variables} in Section \ref{section: figures}.
It is checked that (\ref{equation of the motion 1}-\ref{equation of the motion 2}) admits a unique solution
given almost any initial condition w.r.t. the measure $\mathsf G^0$ defined in Subsection \ref{subsec: Invariant measure} below.
Indeed, looking at figure \ref{figure: projection sur un plan a deux variables}, 
the deterministic solution (when none of the Poisson clock rings) can be constructed at hand for a fixed finite time interval.
The general case can then be obtained, noting that the Poisson processes have almost surely finitely many jumps in any time interval. 

Observe that the condition $p_{k+1}(t_-) < p_k(t_-)$ in the definition of $\mathrm N_k^{\mathrm c}$ is such that particles may move from $\Phi - \Phi^\epsilon$
to $\Phi^\epsilon$, but may not escape from $\Phi^{\epsilon}$.
The definition of the dynamics outside $\Phi^\epsilon$ is absolutely irrelevant for the statement of Theorem \ref{theorem: principal result}, 
but our choice turns out to be convenient for its proof.

\subsection{Evolution semi-group and generator}
Let $\epsilon \ge 0$.
For $t \ge 0$, the operator $\mathcal P^{\epsilon,t}$ acts on bounded functions $u$ on $\Phi$ according to the formula 
\begin{equation*}
\mathcal P^{\epsilon,t} u (x) \; = \; \Mean (u \circ X^\epsilon (x,t)).
\end{equation*}
Its adjoint $\mathcal P^{\epsilon *,t}$ acts on measures on $\Phi$, and is defined by means of the relation
\begin{equation*}
\mathcal P^{\epsilon *,t} \mu (u) \; = \; \mu (\mathcal P^{\epsilon,t} u).
\end{equation*} 

Let now $u$ be some smooth function on $\Phi$, and let $v \equiv v (x,t) := \chi_{\Phi^\epsilon}(x) \multiplication \mathcal P^t u (x) $.
The evolution equations imply that (\ref{equation of the motion 1}-\ref{equation of the motion 2}) that $v$ satisfy the boundary conditions 
\begin{align}
&v(\dots, q_j, \dots ,p_j, \dots ,t) = v(\dots, q_j, \dots ,-p_j, \dots ,t) \;\;\text{if}\;\; q_j = 0 \text{ or } q_j=0, \label{CB 1}\\
&v(\dots, q_j,q_{j+1},\dots, p_j,p_{j+1},\dots,t) = v(\dots, q_j,q_{j+1},\dots, p_{j+1},p_j,\dots,t) \;\;\text{if}\;\; q_{j+1} = q_j - 1 + \epsilon, \label{CB 2}
\end{align} 
the first of these relations being satisfied for $1 \le j \le N$, and the second for $1 \le j \le N-1$.
Moreover, it solves the following differential equation in the interior of $\Phi$:
\begin{equation}\label{generator}
\partial_t v 
\; =\;
\sum_{k=1}^N p_k \partial_{q_k} v
+ 
\lambda
\sum_{k=1}^N  \big( v(\dots , -p_k, \dots) - v (\dots , p_k , \dots) \big).
\end{equation}

\subsection{Invariant measure}\label{subsec: Invariant measure}

The probability measure $\mathsf G^{\epsilon}$ is defined as the uniform probability measure on $\Phi^\epsilon$:
\begin{equation*}
\mathsf G^{\epsilon} (u)
\equiv 
\int_\Phi u (x) \, \mathsf G^{\epsilon} (\dd x)
=
\frac{1}{|\PP^N|}\sum_{\mathbf p \in \PP^N} \frac{1}{|\QQ^\epsilon|} \int_{\QQ^\epsilon} u(\mathbf q,\mathbf p) \, \dd \mathbf q,
\end{equation*}
where 
\begin{equation*}
|\PP^N| = (2N')^N
\quad\text{and}\quad
|\QQ^\epsilon| = \int_{\QQ^\epsilon} \dd \mathbf q.
\end{equation*}
When $\epsilon = 0$, we will just write  $\mathsf G$ instead of $\mathsf G^{0}$.
One checks using \eqref{generator} together with the boundary conditions (\ref{CB 1}-\ref{CB 2}) that the measure $\mathsf G^{\epsilon}$ is invariant under the dynamics, 
meaning that $\mathcal P^{\epsilon *,t}\mathsf G^{\epsilon} = \mathsf G^{\epsilon}$ for every $t\ge 0$.  
Since energy is conserved, this is not the only invariant measure. 
In fact $\mathsf G^{\epsilon}$ corresponds to a Gibbs measure at infinite temperature.

\section{Proof of Theorem \ref{theorem: principal result}}\label{Section: proof of the theorem}
Let $0 < \epsilon_0 < 1/2$.
One first shows that the family $(\mathcal E^\epsilon)_{\epsilon \le \epsilon_0}$ is tight in $\mathcal D([0,1])$. 
One then proves that the law of $\mathcal E^\epsilon (t)$ converges weakly for every $t\in [0,1]$ to the law $\overline\Proba (\argument ,t)$ 
characterized by (\ref{limit evolution measures1}-\ref{limit evolution measures2}). 
One finally establishes that the limit process $\mathcal E$ is markovian.

\subsection{Tightness}

We prove here that the family $(\mathcal E^\epsilon)_{\epsilon \le \epsilon_0}$ is tight in $\mathcal D \big( [0,1],\{ \mathsf e_1, \dots , \mathsf e_{N'} \}^N \big)$, meaning that
for every $\delta >0$, there exists a compact set $K_\delta \subset\mathcal D \big( [0,1],\{ \mathsf e_1, \dots , \mathsf e_{N'} \}^N \big)$
such that $\Proba^\epsilon (K_\delta) \ge 1 - \delta$ for every $\epsilon\in ]0,\epsilon_0]$, where $\Proba^\epsilon$ is the law of $\mathcal E^\epsilon$.
For $n\ge 1$, let $K_n$ be the set of functions in $\mathcal D \big( [0,1],\{ \mathsf e_1, \dots , \mathsf e_{N'} \}^N \big)$ having at most $n$ jumps. 
Since $\{ \mathsf e_1, \dots , \mathsf e_{N'} \}^N $ is finite, the sets $K_n$ are compacts. 
Let now $\delta \in ]0,1[$, and let us show that there exists $n(\delta) \ge 1$ such that $\Proba^\epsilon (K_{n(\delta)})\ge 1 - \delta$ for every $\epsilon\in ]0,\epsilon_0]$.

The energy of the particles can only change due to a collision with a neighbor, and so it suffices to establish that 
there exists a sequence $(c_n)_{n\ge 1}$, with $c_n \rightarrow 0$ as $n\rightarrow \infty$, such that, for every $\epsilon \in ]0,\epsilon_0]$ and $1\le k \le N-1$, 
\begin{equation}\label{tightness 1}
\Proba_\mu \Big( \int_0^{\epsilon^{-1}} \dd \mathrm  N^{\mathrm c}_k (s) \; \ge \; n \Big) \; \le \; c_n, 
\end{equation}
where $\mu\in \mathcal S^\epsilon$ is the law of $X^\epsilon (\argument,0)$.

One would like to replace the integrand in this expression by an explicit function on the Markov chain $X^\epsilon$.
If the dynamics were deterministic (no Poisson processes), one should have
\begin{equation}\label{tightness 2}
\int_0^{\epsilon^{-1}} \dd \mathrm  N^{\mathrm c}_k (s) 
\; \le \;
\mathrm C  \int_0^{\epsilon^{-1}} \eta^{-1} \multiplication \chi_{\mathrm Y_k(\epsilon,\eta)} \circ X^\epsilon_s \, \dd s 
\end{equation}
with $\eta>0$ as small as we want, and 
\begin{equation*}
\mathrm Y_k (\epsilon,\eta) \; := \; \{ x \in \Phi : |q_{k+1} - q_k + 1 - \epsilon | \le \eta  \}.
\end{equation*}
But, for $\epsilon>0$ one finds $\eta \equiv \eta (\epsilon)$ small enough such that the event
\begin{equation*}
\Delta(\epsilon, \eta) \; := \; \big\{ \omega : \max_{t\in [0,\epsilon^{-1}]} 
\big(\mathrm N_k (t+ \eta) - \mathrm N_k(t)\big) + \big(\mathrm N_{k+1} (t+ \eta) - \mathrm N_{k+1}(t)\big) \; \le \; 1\big\}
\end{equation*}
has a probability as close to 1 as one wishes. 

Therefore, taking $\eta$ small enough for a given $\epsilon$, one may replace \eqref{tightness 1} by
\begin{equation}
\Proba_\mu \Big( \int_0^{\epsilon^{-1}} \dd \mathrm  N^{\mathrm c}_k (s) \; \ge \; n \;\Big|\; \Delta(\epsilon, \eta) \Big) \; \le \; c_n. 
\end{equation}
Doing so, inequality \eqref{tightness 2} may now be applied (with a larger constant) even in the presence of the Poisson processes, and,  by Markov's inequality, one gets
\begin{align*}
\Proba_\mu \Big( \int_0^{\epsilon^{-1}} \dd \mathrm  N^{\mathrm c}_k (s) \; \ge \; n \;\Big|\; \Delta(\epsilon, \eta) \Big) 
\; \le &\;
\mathrm C\, n^{-1}\multiplication \Mean_\mu  \int_0^{\epsilon^{-1}} \eta^{-1}\multiplication \chi_{\mathrm Y_k(\epsilon,\eta)} \circ X^\epsilon_s \, \dd s \\
=&\;
\mathrm C\, n^{-1} \multiplication \int_0^{\epsilon^{-1}} \dd s \int \mathcal P^{\epsilon, t } (\eta^{-1}\multiplication\chi_{\mathrm Y_k(\epsilon,\eta)} ) \, \dd \mu \\
\le&\;
\mathrm C\, n^{-1} \epsilon^{-1} \multiplication  \mathsf G^{\epsilon} (\eta^{-1}\multiplication \chi_{\mathrm Y_k(\epsilon,\eta)} ) 
\;\le\;
\mathrm C \, n^{-1},
\end{align*} 
where one has used the fact that $\mu \le \mathrm C\, \mathsf G^{\epsilon}$ to get the second inequality, 
and the fact that $\mathsf G^{\epsilon} (\eta^{-1}\multiplication \chi_{\mathrm Y_k(\epsilon,\eta)} )  = \mathcal O (\epsilon)$ for any value of $\eta \in ]0,\epsilon]$ to get the last one. 
$\square$

\subsection{Convergence at a given time}
We show here that the law of $\mathcal E^\epsilon (t)$ converges weakly for every $t\in [0,1]$ to the law $\overline\Proba (\argument ,t)$, 
characterized by (\ref{limit evolution measures1}-\ref{limit evolution measures2}). 
Observe that, for given $\epsilon >0$, there is an obvious identification between functions on $\{\mathsf e_1, \dots \mathsf e_{N'}\}^N$ and measures in $\mathcal S^\epsilon$, 
that will be used in the sequel.

Let $\Proba^\epsilon (\argument , t)$ be the law of $X^\epsilon (t)$ (so \emph{not} the law of $\mathcal E^\epsilon (t)$ as it was in the previous subsection).
We actually will prove that,
for any $t\in [0,1]$,
there exists a probability measure $\overline\Proba (\argument , t)\in \mathcal S^0$, such that $\Proba^\epsilon (\argument, t)$ converges weakly to $\overline\Proba (\argument , t)$ as $\epsilon \rightarrow 0$, 
and then that
$\overline\Proba (\argument, t)$ solves (\ref{limit evolution measures1}-\ref{limit evolution measures2}).
This will imply our result.
Indeed, the distribution of $\mathcal E^\epsilon (t)$ may be written as 
$\int_\Phi \Proba^\epsilon (\dd x, t | \mathbf e)$, 
and so, as $\epsilon \rightarrow 0$, the distribution of $\mathcal E^\epsilon (t)$ also will converge to $\overline{\Proba}(\argument, t)$.
We will use the following lemma, which proof is deferred to the next section. 

\begin{Lemma}\label{lemma: fundamental technical result}
Let $v$ be a function on $\{ \mathsf e_1,\dots , \mathsf e_{N'}\}^N$
corresponding to a probability measure in $\mathcal S^\epsilon$ for every $\epsilon\in ] 0,\epsilon_0 ]$.
For $\tau >0$ small enough, 
and for $0 < \epsilon < \tau^2$, 
\begin{equation*}
\mathcal P^{*\epsilon,\epsilon^{-1}\tau} \, \Proba
\; = \; 
\Big(1 - \tau \sum_{k=1}^{N-1}\gamma (e_k,e_{k+1})\Big)\,  \Proba 
\; + \;  
\tau \sum_{k=1}^{N-1} \gamma  (e_k,e_{k+1})\,\Proba (\dots , e_{k+1}, e_k,\dots)
\; +\; 
\mathcal O (\tau^2) \, \mu_{\epsilon, \tau},
\end{equation*}
where $\Proba\in \mathcal S^\epsilon$ is the measure determined by $v$, 
and where $\mu_{\epsilon, \tau}$ is a measure on $\Phi$ such that $\| \mu_{\epsilon, \tau} \|_1 \le 1$.
\end{Lemma}
\noindent

Let now $t \in [0,1]$.
Let $\epsilon \in ]0,\epsilon_0[$, and let $n_\epsilon$ be the largest integer such that $\tau := t/n_\epsilon > \epsilon^{1/2}$.
One has $\Proba^\epsilon (\argument, t) = \mathcal P^{\epsilon*,\epsilon^{-1}t} \Proba^\epsilon (\argument, 0)$.
Lemma \ref{lemma: fundamental technical result} applies if $\epsilon$ is small enough, and one has
\begin{equation*}
\mathcal P^{\epsilon*,\epsilon^{-1}t} \Proba (\argument, 0)
\; = \;
\mathcal P^{\epsilon*,\epsilon^{-1} n_\epsilon\tau} \Proba (\argument, 0)
\; = \;
\mathcal P^{\epsilon*,\epsilon^{-1}(n_\epsilon-1)\tau}
\big( \Proba_1 +  \mathcal O (n_\epsilon^{-2}) \, \mu_{1, \epsilon, \tau} \big),
\end{equation*}
where $\Proba_1\in \mathcal S^\epsilon$ is the measure 
\begin{equation*}
\Proba_1 \; := \; \Big(1 - \tau \sum_{k=1}^{N-1}\gamma (e_k,e_{k+1})\Big)\,  \Proba^\epsilon (\argument, 0)
\; + \;  
\tau \sum_{k=1}^{N-1} \gamma  (e_k,e_{k+1})\,\Proba^\epsilon (\dots , e_{k+1}, e_k,\dots,0), 
\end{equation*}
and $\mu_{1, \epsilon, \tau}$ a measure such that $\| \mu_{1, \epsilon, \tau} \|_1 \le 1$.
Observe that $\Proba_1$ is a probability measure if $\tau$ is small enough, what we assume.
Since $\| \mathcal P^{\epsilon*,s}\mu \|_1 \le \| \mu \|_1$ for every measure $\mu$ and every $s\ge 0$, one may iterate this and get 
\begin{equation*}
\mathcal P^{\epsilon*,\epsilon^{-1}t} \Proba^\epsilon (\argument, 0)
\; = \;
\Proba_{n_\epsilon}
+
\mathcal O (n_{\epsilon}^{-2}) \sum_{k=1}^{n_\epsilon} \tilde \mu_{k, \epsilon, \tau},
\end{equation*}
with $\Proba_{n_\epsilon}$ a probability measure in $\mathcal S^\epsilon$, 
and $\tilde \mu_{k, \epsilon, \tau}$ measures such that $\|\tilde \mu_{k, \epsilon, \tau} \|_1 \le 1$.

Let us first show the existence of a probability measure $\overline{\Proba}(\argument,t)\in \mathcal S^0$ 
in the closure of $(\Proba_{n_\epsilon})_{\epsilon \le \epsilon_0}$ for the $\| \argument\|_1$-norm.
For this, let $v_{n_\epsilon}$ be the function on $\{ \mathsf e_1,\dots , \mathsf e_{N'}\}^N$ determining the measure $\Proba_{n_\epsilon}\in\mathcal S^\epsilon$, 
and let then $\tilde\Proba_{n_\epsilon}$ be the unique measure in $\mathcal S^0$ determined by $v_{n_\epsilon}$. 
Since $\| \Proba_{n_\epsilon} - \tilde\Proba_{n_\epsilon} \|_1 \le \mathrm C\, \epsilon^2 $, 
a point in the closure of $(\tilde\Proba_{n_\epsilon})_{\epsilon \le \epsilon_0}$
is also in the closure of $(\Proba_{n_\epsilon})_{\epsilon \le \epsilon_0}$ and is a probability measure.
But, since $\mathcal S^0$ is a finite-dimensional space, and since $(\tilde\Proba_{n_\epsilon})_{\epsilon \le \epsilon_0}$ is a bounded set for the $\| \argument\|_1$-norm, 
the announced measure  $\overline{\Proba}(\argument, t)$ indeed exists.

Let us next show that $\overline\Proba(\argument, t)$ solves (\ref{limit evolution measures1}-\ref{limit evolution measures2}).
Up to a subsequence, one can write 
\begin{equation}\label{convergence forte de la mesure de proba}
\overline\Proba(\argument, t) \; = \; \lim_{\epsilon \rightarrow 0} \mathcal P^{\epsilon*,\epsilon^{-1}t}\Proba (\argument, 0)
\end{equation}
for the $\| \argument\|_1$-norm, and so
\begin{eqnarray*}
\lim_{\tau\rightarrow 0} \frac{\overline \Proba(\argument,{t+\tau}) - \overline \Proba(\argument, t)}{\tau}
&=&
\lim_{\tau\rightarrow 0} \frac{1}{\tau} \lim_{\epsilon \rightarrow 0} 
\big( \mathcal P^{\epsilon*,\epsilon^{-1}\tau} - \mathrm{Id} \big) \mathcal P^{\epsilon*,\epsilon^{-1}t}\Proba (\argument , 0) \\
&=& 
\lim_{\tau\rightarrow 0} \frac{1}{\tau} \Big(
\lim_{\epsilon \rightarrow 0} \big( \mathcal P^{\epsilon*,\epsilon^{-1}\tau} - \mathrm{Id} \big) \overline\Proba (\argument, t) \Big) \\
&&+
 \lim_{\tau\rightarrow 0} \frac{1}{\tau} \Big(
\lim_{\epsilon \rightarrow 0} \big( \mathcal P^{\epsilon*,\epsilon^{-1}\tau} - \mathrm{Id} \big) 
\big(\mathcal P^{\epsilon*,\epsilon^{-1}t}\Proba (\argument, 0) - \overline\Proba (\argument, t) \big)
\Big).
\end{eqnarray*}
The second term in the right hand side of this equation is zero since 
\begin{equation*}
\big\| \big( \mathcal P^{\epsilon*,\epsilon^{-1}\tau} - \mathrm{Id} \big) \big(\mathcal P^{\epsilon*,\epsilon^{-1}t}\Proba (\argument, 0) - \overline\Proba (\argument, t) \big)  \big\|_1 
\le
2 \, \big\|\mathcal P^{\epsilon*,\epsilon^{-1}t}\Proba (\argument, 0) - \overline\Proba (\argument, t)  \big\|_1 
\rightarrow 0
\end{equation*}
as $\epsilon\rightarrow 0$.
By Lemma \ref{lemma: fundamental technical result} instead, the first term gives 
\begin{equation*}
\lim_{\tau\rightarrow 0} \frac{1}{\tau} \Big(
\lim_{\epsilon \rightarrow 0} \big(  \mathcal P^{\epsilon*,\epsilon^{-1}\tau} - \mathrm{Id} \big) \overline\Proba (\argument, t) \Big) 
=
\sum_{k=1}^{N-1}\gamma (e_k,e_{k+1})
\Big( 
\overline\Proba (\dots , e_{k+1},e_k,\dots,t) - \overline\Proba (\dots , e_k, e_{k+1},\dots,t)
\Big),
\end{equation*}
which establishes the result. $\square$

\subsection{$\mathcal{E}$ is a Markov process}
We show that, for any $0= t_0 < t_1 < \dots < t_n \le 1$ and any $\mathbf e_0 , \mathbf e_1 ,\dots , \mathbf e_n \in \{ \mathsf e_1, \dots , \mathsf e_{N'} \}$, for $n\ge 1$,
\begin{equation*}
\Proba \big(\mathcal E_{t_n} = \mathbf e_n \big| \mathcal E_{t_{n-1}} = \mathbf e_{n-1}, \dots , \mathcal E_{0} = \mathbf e_0 \big)
\; = \; 
\Proba \big(\mathcal E_{t_n - t_{n-1}} = \mathbf e_n \big| \mathcal E_{0} = \mathbf e_{n-1}\big).
\end{equation*}
This is immediate for $n=1$. We show the case $n=2$, the generalization being rather straightforward. 
For $\epsilon >0$, since $X^{\epsilon}$ is a Markov process, one may write
\begin{equation*}
\Proba \big(\mathcal E^\epsilon_{t_2} = \mathbf e_2 \big| \mathcal E^\epsilon_{t_1} = \mathbf e_1 , \mathcal E^\epsilon_{0} = \mathbf e_0  \big)
\; = \; 
\Proba_{\mu^\epsilon} \big( \mathcal E^{\epsilon}_{t_2 - t_1} = \mathbf e_2 \big)
\end{equation*}
where $\mu^\epsilon$ is an initial measure on $\Phi$ given by
\begin{equation*}
\mu^\epsilon \; = \; \mathcal P^{\epsilon *,\epsilon^{-1} t_1} \mu_0 (\argument | \mathbf e = \mathbf e_1)
\quad \text{with} \quad
\mu_0 \; = \; \mathsf G^\epsilon (\argument | \mathbf e  = \mathbf e_0).
\end{equation*}
Our claim will now be established if we show that $\| \mu^\epsilon - \mathsf G^\epsilon (\argument | \mathbf e = \mathbf e_1) \|_1 \rightarrow 0$ as $\epsilon \rightarrow 0$.
But this follows from the facts that 
 $\|  \mathcal P^{\epsilon *,\epsilon^{-1} t_1} \mu_0 -  \overline{\Proba}(\argument, t_1 )\|_1 \rightarrow 0$ as $\epsilon \rightarrow 0$ 
by \eqref{convergence forte de la mesure de proba}, 
that $\overline{\Proba}(\argument, t_1| \mathbf e= \mathbf e_1) = \mathsf G (\argument | \mathbf e= \mathbf e_1)$ since $\overline \Proba (\argument ,t_1)\in \mathcal S^0$,
and that $\|\mathsf G - \mathsf G^\epsilon \|_1  = \mathcal O (\epsilon^2)$.
$\square$

\section{Proof of Lemma \ref{lemma: fundamental technical result}}\label{Section: proof of the main lemma}

The proof of Lemma \ref{lemma: fundamental technical result} is conceptually very easy.
We start by sketching the method in Subsection \ref{subsection: method}. 
In Subsection \ref{subsection: intermediate lemma to the proof of the lemma}, some crucial elementary facts are collected. 
We finally prove Lemma \ref{lemma: fundamental technical result} in Subsection \ref{subsection: proof of the lemma main technical result}.

\subsection{Method}\label{subsection: method}
Since $\Proba \in \mathcal S^\epsilon$, this measure is invariant under the uncoupled dynamics, up to an error of order $\epsilon^2$ for the $\| \argument \|_1$-norm.
We recall that $\epsilon < \tau^2$ by hypothesis.   
Writing the result of Lemma \ref{lemma: fundamental technical result} as 
\begin{equation}\label{method 1}
\mathcal P^{\epsilon *,\epsilon^{-1}\tau} \Proba -  \mathcal P^{0*,\epsilon^{-1}\tau}\Proba = \tau \, (\dots) + \mathcal O (\tau^2),
\end{equation}
one sees that one needs to evaluate the difference between the evolution of the coupled and the uncoupled dynamics in the first order in $\tau$.
The following identity, known as Duhamel's formula, is suited for this: 
if $S$ and $T$ are two bounded operators on some space, one has 
\begin{equation}\label{Duhamel 1er ordre}
T^n  - S^n 
\;=\; 
\sum_{k=1}^n T^{k-1}(T-S)S^{n-k}. 
\end{equation}
Iterating this formula to replace $T^{k-1}$ by $S^{k-1}$, one finds 
\begin{equation}\label{method 2}
T^n  - S^n 
\;=\; 
\sum_{k=1}^n S^{k-1}(T-S)S^{n-k}
+
\sum_{k=1}^n \sum_{j=1}^{k-1} T^{j-1} (T-S) S^{k-1-j} (T-S) S^{n-k}.
\end{equation}

Let now $\eta$ be some small time step (see Subsection \ref{subsection: notations}), 
let $S = \mathcal P^{0*,\eta}$, let $T = \mathcal P^{\epsilon *,\eta}$, and let $n$ be such that $\epsilon^{-1}\tau=n\eta$.
At this point, it could have seemed more natural to work directly in a continuous-time setting. 
In that case, the sums appearing in \eqref{Duhamel 1er ordre} and \eqref{method 2} becomes time integral, 
whereas $T-S$ should become the part of the generator corresponding to a collision between particles.
The problem is that this term is hard to make explicit, since it is only present through the boundary conditions (\ref{CB 1}-\ref{CB 2}), 
and that is why we decided to divide time in arbitrary small intervals.  

All what needs to be done now is to show that 
the first term in the right hand side of \eqref{method 2} coincides with the first term in the right hand side of \eqref{method 1} up to an error of order $\epsilon$, 
and that the second term is $\mathcal O (\tau^2)$.  
The identification of the first terms is not hard 
since one has a good control on the uncoupled dynamics (see Section \ref{section: uncoupled dynamics}), i.e. on $S= \mathcal P^{0*,\eta}$, 
and since one can estimate the difference $T-S = \mathcal P^{\epsilon *,\eta} - \mathcal P^{0*,\eta}$ for $\eta$ small enough, 
as explained in Subsection \ref{subsection: intermediate lemma to the proof of the lemma} below.

The second term represents recollisions. 
One needs to see that, in a macroscopic time of order $\tau$, the probability of two collisions or more for the same particle is $\mathcal O(\tau^2)$, 
as it should be if they were happening at random. 
This turns out to be a bit heavy to show, mainly due to the persistence of a deterministic trajectory for microscopic time intervals.
The diophantine condition on the velocities has here to be used.

\subsection{Crucial elementary estimates}\label{subsection: intermediate lemma to the proof of the lemma}
Let $\epsilon\ge 0$ and let $\epsilon > \eta > 0$ be a small time interval.
Let us define some subsets of $\Phi$.
For this, let 
\begin{equation*}
\mathsf p
\; := \; 
2 \max \big\{ \sqrt{2\mathsf e_1}, \dots ,  \sqrt{2\mathsf e_{N'}} \big\}.
\end{equation*}
Then, for $1 \le k \le N-1$, 
\begin{align}
\mathrm Z_k
\; = &\; 
\{ x \in \Phi : |q_{k+1} - q_k + 1 - \epsilon | \le \mathsf p \eta \}, 
\label{def of Z k} \\
\mathrm{Coll}_k
\; =&\; 
\{ x\in \Phi : 
\tilde{\mathrm N}^{\mathrm c}_k (x,\eta) - \tilde{\mathrm N}^{\mathrm c}_k (x,0) = 1 \}. 
\label{def of Coll k}
\end{align}
Here $\tilde{\mathrm N}^{\mathrm c}_k (x,s)$ is the value at time $s$ of the process $\mathrm N_k^\mathrm c$ defined in Subsection \ref{subsection: def of the dynamics}
for a deterministic trajectory starting at $x$.
We recall that the use of a tilde always refers to the deterministic dynamics (see Subsection \ref{subsection: notations}).
One then sets
\begin{equation*}
\mathrm Z \; = \; \bigcup_{k=1}^{N-1} \mathrm Z_k , 
\qquad
\mathrm{Coll} \; = \; \bigcup_{k=1}^{N-1} \mathrm{Coll}_k.
\end{equation*}

First, for $1 \le k \le N-1$, the characteristic function of the set $\mathrm Z_k$ depends only on the variables $q_k$ and $q_{k+1}$. 
A computation yields 
\begin{equation}\label{size of Zk}
\mathrm C\,  \epsilon  \eta 
\; \le \; 
\int_{[0,1]^2} \chi_{\mathrm Z_k} \, \dd q_k \dd q_{k+1}
\; \le \; 
\int_{\QQ} \chi_{\mathrm Z} \, \dd \mathbf q
\; \le \; 
\mathrm C' \,  \epsilon  \eta ,
\end{equation} 
and, for $1 \le j\ne k \le N-1$, 
\begin{equation}\label{intersection des Z}
\int_{\QQ} \chi_{\mathrm Z_j} \multiplication \chi_{\mathrm Z_k} \, \dd \mathbf{q} \; = \; \mathcal O (\epsilon \eta)^2.
\end{equation}

Second, for $1 \le k \le N-1$, let us define
\begin{align*}
\mathrm W_k 
\; =  & \; 
\{ x\in \Phi : 1 - q_k \le \mathsf p \eta \text{ and } |q_{k+1} - q_k + 1 - \epsilon | \le \mathsf p \eta \}, \\
\mathrm W'_k
\; =  & \; 
\{ x\in \Phi : q_{k+1} \le \mathsf p \eta \text{ and } |q_{k+1} - q_k + 1 - \epsilon | \le \mathsf p \eta \}.
\end{align*}
The deterministic dynamics is such that 
\begin{align}
\mathrm{Coll} 
\; \subset &\;
\{ x \in \Phi: p_k > p_{k+1} \text{ and } 0 \le q_{k+1} - q_k + 1 - \epsilon \le (p_k - p_{k+1})\eta  \} \cup \mathrm W_k \cup \mathrm W'_k ,  \label{Coll estim inf}\\
\mathrm{Coll} 
\; \supset &\;
\{ x \in \Phi: p_k > p_{k+1} \text{ and } 0 \le q_{k+1} - q_k + 1 - \epsilon \le (p_k - p_{k+1})\eta \}  -  \big( \mathrm W_k \cup \mathrm W'_k \big).\label{Coll estim sup}
\end{align}
So $\mathrm{Coll}_k \subset \mathrm Z_k$ and thus $\mathrm{Coll} \subset \mathrm Z$, and 
the characteristic function of $\mathrm{Coll}_k$ only depends on the variables $x_k$ and $x_{k+1}$.
A computation based on (\ref{Coll estim inf}-\ref{Coll estim sup}) gives
\begin{equation}\label{size of Collk}
\int_\QQ \chi_{\mathrm{Coll}_k} \, \dd q_k \dd q_{k+1}
\; = \; 
\chi_{\R_+}(p_k - p_{k+1}) \multiplication (p_k - p_{k+1}) \eta \epsilon
+
\mathcal O (\eta^2),
\end{equation}
for any values of the impulsions.
It follows moreover from \eqref{intersection des Z} that, for $1 \le j\ne k \le N-1$,
\begin{equation}\label{intersection des Coll}
\int \chi_{\mathrm{Coll}_j} \multiplication \chi_{\mathrm{Coll}_k} \, \dd \mathbf q \; = \; \mathcal O (\epsilon \eta)^2
\end{equation}
for any values of the impulsions.

Let us finally estimate the difference between the evolution of the coupled and uncoupled dynamics over a time interval $\eta$.
Let us define the operator
\begin{equation}\label{def of L}
L \; := \; \frac{1}{\eta} (\mathcal P^{\epsilon, \eta} - \mathcal P^{0,\eta}).
\end{equation}
One has
\begin{equation}\label{def of L tilde}
\tilde L 
\; := \; 
\frac{1}{\eta}(\tilde{\mathcal P}^{\epsilon,\eta} - \tilde{\mathcal P}^{0,\eta}) 
\; = \; 
\chi_{\mathrm{Coll}} \multiplication \frac{1}{\eta}(\tilde{\mathcal P}^{\epsilon,\eta} - \tilde{\mathcal P}^{0,\eta})
\end{equation}
and
\begin{Lemma}\label{lemma: L and tilde L}
There exists a constant $\mathrm C <+\infty$ such that, for all $\eta >0$ small enough, and for all $u\in \Lp^\infty(\Phi)$,
\begin{equation*}
L u 
=
\tilde L u + \chi_{\mathrm Z} \, u'
\end{equation*}
where $u'\in \Lp^{\infty}(\Phi)$ is such that $\| u' \|_\infty \le \mathrm C \| u \|_\infty$.
\end{Lemma}

\Proof
Let $x\in \Phi$.
If $x\notin \mathrm Z$, then $X^{\epsilon}(x,\eta) = X^{0}(x,\eta)$ for every realization of the Poisson processes. 
Therefore
\begin{equation*}
L u = \chi_{\mathrm Z}\multiplication  Lu.
\end{equation*}
Next, the probability that one of the Poisson processes have a jump in a time interval of length $\eta$ is itself $\mathcal O (\eta)$
so that 
\begin{equation*}
\mathcal P^{\epsilon,\eta} u (x) 
= \Mean (u \circ X^{\epsilon}(x,\eta)) 
= u\circ \tilde X^{\epsilon}(x,\eta) + \mathcal O (\eta \| u \|_\infty) 
= \tilde{\mathcal P}^{\epsilon,\eta} u (x) + \mathcal O (\eta \| u \|_\infty).
\end{equation*}
This last formula still holds also in particular for $\epsilon = 0$, and therefore
$L u = \frac{1}{\eta} \tilde L u + \chi_{\mathrm Z} u'$ with $\| u'\|_\infty \le \mathrm C\, \|  u\|_\infty$ for some $\mathrm C < + \infty$. $\square$

\subsection{Proof of Lemma \ref{lemma: fundamental technical result}}\label{subsection: proof of the lemma main technical result}

\paragraph{1. Preliminary simplifications.}
We will give the proof in the case $N' = N$, indicating at the very end the small needed adaptations to bring for dealing with the case $N' < N$.
The space $\mathcal S^\epsilon$ is generated by functions of the form 
\begin{equation*}
\chi_{\mathsf e_{k_1}, \dots , \mathsf e_{k_N}} (e_1,\dots ,e_N ) \multiplication \chi_{\QQ^\epsilon} (\mathbf q),
\end{equation*}
where $k_1, \dots ,k_N$ is a permutation of $1 , \dots , N$.
By linearity, it suffices to consider the case where $\mathsf P$ is proportional to a function of that form. 
Moreover, to simplify notations, we will assume that $(k_1, \dots , k_N) = (1 , \dots , N)$.
Let then 
\begin{equation}\label{def of P prime}
\mathsf P' 
\; := \; \frac{1}{2^N} \multiplication \chi_{\mathsf e_1, \dots , \mathsf e_N} (e_1, \dots , e_N) \multiplication \chi_\QQ (\mathbf q) 
\; = \; \frac{1}{2^N} \multiplication \chi_{\mathsf e_1, \dots , \mathsf e_N} (e_1, \dots , e_N).
\end{equation}
This measure
is close to $\Proba$ when $\epsilon \rightarrow 0$: 
\begin{equation*}
\| \Proba - \Proba' \|_1 
\; = \; 
\mathcal O (\epsilon^2).
\end{equation*}
Because $\| \mathcal P^{\epsilon *,t }\mu \|_1 \le \| \mu\|_1$ for every $\mu$ and every $t\ge 0$, 
one may therefore show the result with $\Proba'$ instead of $\Proba$.
This will be advantageous since $\Proba'$ is invariant under the uncoupled dynamics.

\paragraph{2. Applying Duhamel's formula.}
Let $\tau >0$, let $\epsilon\in ]0,\tau^2[$, and let $\eta\in ]\epsilon^4,2\epsilon^4[$ be such that, for some integer $n\ge 1$, one has
\begin{equation}
\epsilon^{-1}\tau = \eta n.
\end{equation}
Applying Duhamel's formula \eqref{Duhamel 1er ordre} and noting that  $\mathsf P^{0*,t} \Proba' = \Proba'$ for every $t\ge 0$, one gets
\begin{equation*}
\mathcal P^{\epsilon*,\epsilon^{-1}\tau} \Proba' = \Proba' + \eta \sum_{k=1}^n \mathcal P^{\epsilon *,(k-1)\eta}L^*\Proba',
\end{equation*}
with $L$ the operator defined in \eqref{def of L}.
One would like to replace $L^*$ by $\tilde L^*$, with $\tilde L$ defined in \eqref{def of L tilde}. 
One has 
\begin{align*}
\| \mathsf P^{\epsilon*,(k-1)\eta} (L^{*} - \tilde L^{*}) \Proba'  \|_1
\le& 
\; \|  (L^{*} - \tilde L^{*})  \Proba'  \|_1 
=
\sup_{u:\| u \|_\infty \le 1} \Proba' \big( (L - \tilde L)  u \big)\\
\le&
\; 
\sup_{u:\| u \|_\infty \le 1} \; 
\max_{\mathbf p\in\PP^N} 
\int_\QQ \big| (L - \tilde L)  u (\mathbf q,\mathbf p) \big| \, \dd \mathbf q
\; = \; 
\mathcal O (\epsilon \eta).
\end{align*}
Here the inequality follows from the fact that $\Proba'$ is a probability measure independent of $\mathbf q$, 
whereas the last equality is a consequence of Lemma \ref{lemma: L and tilde L} and \eqref{size of Zk}.
Therefore 
\begin{equation}\label{B n p nouvelle mouture}
\mathcal P^{\epsilon*,\epsilon^{-1}\tau} \Proba' = \Proba' + \eta \sum_{k=1}^n \mathcal P^{\epsilon *,(k-1)\eta} \tilde L^*\Proba'
+
\mu_0
\end{equation}
where $\mu_0$ is a measure such that
\begin{equation}\label{borne sur mu 0}
\|\mu_0 \|_1 
= 
\mathcal O (\eta n \multiplication \epsilon \eta) 
= 
\mathcal O (\tau^2).
\end{equation}
One then uses Duhamel's formula once more to replace $\mathcal P^{\epsilon *,(k-1)\eta}$ by $\mathcal P^{0 *,(k-1)\eta}$ in \eqref{B n p nouvelle mouture}:
\begin{align*}
\mathcal P^{\epsilon*,\epsilon^{-1}\tau} \Proba'  
&=\; \Proba' + \eta \sum_{k=1}^n \mathcal P^{0 *,(k-1)\eta} \tilde L^*\Proba' 
  + \eta^2 \sum_{k=1}^n \sum_{j=1}^{k-1} \mathcal P^{\epsilon*,(j-1)\eta} L^* \mathcal P^{0*, (k-1 -j)\eta} \tilde L^*\Proba' + \mu_0 \\
&:=\; \Proba' + \mu_1 + \mu_2 + \mu_0.
\end{align*}
It is not clear at this point that we can replace the remaining operator $L^*$ by $\tilde L^*$ up to a negligible error, 
and so we directly proceed now to the analysis of $\mu_1$ and $\mu_2$. 
In a first step we will show that $\mu_1$ is given by \eqref{expression finale de mu1} below, 
whereas in a second step, one will prove that $\| \mu_2\|_1 = \mathcal O (\tau^2)$. 
Thanks to the bound \eqref{borne sur mu 0} on $\|\mu_0\|_1$, this will finish the proof.

\paragraph{3. Analyzing $\mu_1$.}
We here show that $\mu_1$ can be expressed as in \eqref{expression finale de mu1} below.
Given a measure $\mu$ on $\Phi$, we write $\mu (1 | \mathbf e)$ the measure in $\mathcal S^0$ explicitly given by
\begin{equation}\label{def projection sur S 0}
\mu (1 | \mathbf e) \; = \; \frac{1}{2^N} \sum_{p_1 = \pm \sqrt{2 e_1}} \dots \sum_{p_N = \pm \sqrt {2 e_n}} \, \int_\QQ \mu (\dd \mathbf q, \mathbf p), 
\end{equation}
where the factor $1/2^N$ appears since we want to see $\mu (1 | \mathbf e)$ as a measure on $\Phi$, and not as a measure on $\{\mathsf e_1, \dots ,\mathsf e_N \}^N$.
Because $\mathcal P^{0*,t} \mu (1|\mathsf e) = \mu (1 |\mathsf e)$ for any $\mu$ on $\Phi$ and $t\ge 0$, 
and since $\eta n = \epsilon^{-1}\tau$, one has
\begin{equation}\label{expression for mu 1}
\mu_1 = 
\epsilon^{-1} \tau \multiplication \tilde L^{*} \Proba' (1 |\mathbf e) + 
\eta \sum_{k=1}^n \mathcal P^{0 *,(k-1)\eta}
\big( \tilde L^{*} \Proba' - \tilde L^{*} \Proba' (1 |\mathbf e)  \big).
\end{equation} 

Let us first compute $\tilde L^{*} \Proba' (1 |\mathbf e)$. 
It follows from \eqref{def projection sur S 0} that
\begin{equation*}
\tilde L^{*} \Proba' (1 | \mathsf e) 
\; = \;
\frac{1}{2^N} \tilde L^{*} \Proba' (\chi_{\mathbf e})
\; = \; 
\frac{1}{2^N} \Proba' (\tilde L \chi_{\mathbf e}).
\end{equation*}
Now, from \eqref{def of L tilde} and then \eqref{intersection des Coll},
\begin{align*}
\Proba' (\tilde L \chi_{\mathbf e})
\; = &\;
\frac{1}{\eta}
\Big( \Proba' (\chi_{\mathrm{Coll}} \multiplication \tilde{\mathcal P}^{\epsilon,\eta} \chi_{\mathbf e})
 -
\Proba' (\chi_{\mathrm{Coll}} \multiplication \tilde{\mathcal P}^{0,\eta} \chi_{\mathbf e})
\Big)\\
\; = &\; \frac{1}{\eta} \sum_{k=1}^{N-1} 
\Big( \Proba' (\chi_{\mathrm{Coll}_k} \multiplication \tilde{\mathcal P}^{\epsilon,\eta} \chi_{\mathbf e})
 -
\Proba' (\chi_{\mathrm{Coll}_k} \multiplication \tilde{\mathcal P}^{0,\eta} \chi_{\mathbf e}) \Big)
\; + \; \mathcal O (\eta).
\end{align*}
On the one hand, from the definition \eqref{def of Coll k}, one has 
\begin{equation*}
\tilde {\mathcal P}^{\epsilon,\eta} \chi_{\mathbf e} (x)
\; = \; 
\chi_{\mathbf e} \big( \tilde X^\epsilon (x,\eta) \big) 
\; = \; 
\chi_{\sigma_k \mathbf e } (x)
\quad\text{for}\quad x\in \mathrm{Coll}_k - \bigcup_{j\ne k} \mathrm{Coll}_j, 
\end{equation*}
with
\begin{equation*}
\sigma_k \mathbf e \; = \; (e_1, \dots , e_{k+1},e_k, \dots , e_N).
\end{equation*}
On the other hand, $\tilde{\mathcal P}^{0,\eta}\chi_{\mathbf e} = \chi_{\mathbf e}$ since the uncoupled dynamics preserves the energy of individual particles. 
Therefore, using \eqref{intersection des Coll} once more, 
\begin{equation*}
\Proba' (\tilde L \chi_{\mathbf e})
\; = \; 
\frac{1}{\eta} \sum_{k=1}^{N-1}
\Big(  \Proba' (\chi_{\mathrm{Coll}_k} \multiplication \chi_{\sigma_k \mathbf e})
 -
\Proba' (\chi_{\mathrm{Coll}_k} \multiplication \chi_{\mathbf e})
\Big)
\; + \; \mathcal O (\eta).
\end{equation*}
From the particular form \eqref{def of P prime} of $\Proba'$ and from \eqref{size of Collk}, one obtains
\begin{align*}
\Proba' (\chi_{\mathrm{Coll}_k} \multiplication \chi_{\sigma_k \mathbf e}) 
\; = &\; 
\frac{1}{2^N} \sum_{\mathbf p' \in \PP^N}
\chi_{(\mathsf e_1,\dots ,\mathsf e_N)}(\mathbf p')\multiplication \chi_{\sigma_k \mathbf e}(\mathbf p')\multiplication \int_\QQ \chi_{\mathrm{Coll}_k}(\mathbf q') \, \dd \mathbf q' \\
=&\; 
\frac{\eta\epsilon}{2^N}\sum_{\mathbf p' \in \PP^N}
\chi_{(\mathsf e_1,\dots ,\mathsf e_N)}(\mathbf p')\multiplication \chi_{\sigma_k \mathbf e}(\mathbf p')\multiplication \chi_{\R_+}(p'_k - p'_{k+1})\multiplication (p'_k - p'_{k+1}) 
\;+ \; \mathcal O(\eta^2)\\
=&\;
\frac{\eta\epsilon}{4}\multiplication \chi_{(\mathsf e_1,\dots ,\mathsf e_N)} (\sigma_k \mathbf e)
\sum_{\substack{p_k' = \pm  \sqrt{2e_{k+1}} \\ p_{k+1}' = \pm \sqrt{2 e_k}}}
\chi_{\R_+}(p'_k - p'_{k+1})\multiplication (p'_k - p'_{k+1}) 
\;+ \; \mathcal O(\eta^2)\\
=&\;
\eta \epsilon  \multiplication \chi_{(\mathsf e_1,\dots ,\mathsf e_N)} (\sigma_k \mathbf e) \multiplication  \gamma (e_k,e_{k+1})
\; + \; \mathcal O (\eta^2),
\end{align*}
where $\gamma$ is the function defined by \eqref{def of gamma}.
A similar computation shows that
\begin{equation*}
\Proba' (\chi_{\mathrm{Coll}_k} \multiplication \chi_{\mathbf e}) 
\; = \;
\eta \epsilon  \multiplication \chi_{(\mathsf e_1,\dots ,\mathsf e_N)} (\mathbf e) \multiplication  \gamma (e_k,e_{k+1})
\; + \; \mathcal O (\eta^2),
\end{equation*}
so that one concludes that 
\begin{align}
\tilde L^{*} \Proba' (1 | \mathbf e) 
\; = &\;
\frac{\epsilon}{2^N} \sum_{k=1}^{N-1}
\gamma (e_k,e_{k+1}) \multiplication 
\big( \chi_{\mathsf e_1,\dots, \mathsf e_N} (\sigma_k \mathbf e) -  \chi_{\mathsf e_1, \dots , \mathsf e_N} (\mathbf e) \big) + \mathcal O (\eta) \nonumber \\
\; = &\;
\epsilon
\sum_{k=1}^N
\gamma (e_k,e_{k+1})\multiplication 
\big(  \Proba' (\sigma_k \mathbf e) - \Proba' (\mathbf e) \big) + \mathcal O (\eta).\label{proof of mu 1 estimate obtained from the projection}
\end{align}

We then need to bound the second term in the right hand side of \eqref{expression for mu 1}.
Let us first observe that, 
by the particular form \eqref{def of P prime} of $\Proba'$, 
by the definition \eqref{def of L tilde} of $\tilde L$, 
by the fact that $\mathrm{Coll} \subset \mathrm Z$, and finally by \eqref{size of Zk}, 
\begin{equation}\label{integrale de tilde L}
\| \tilde L^{*} \Proba' \|_1
\; \le \;
\sup_{u:\|u\|_\infty \le 1} \,
\max_{\mathbf p \in \PP^N}\,
\int_\QQ |\tilde L u (\mathbf q , \mathbf p) | \, \dd \mathbf q
\; \le \;
\max_{\mathbf p \in \PP^N}\,
\frac{2}{\eta} \, \int_\QQ \chi_{\mathrm Z} (\mathbf q) \, \dd \mathbf q
\; = \; 
\mathcal O ( \epsilon ).
\end{equation}
The uncoupled dynamics is studied in more details in the next section ; 
it follows from \eqref{integrale de tilde L} and \eqref{uncoupled dynamics result 1} there that 
\begin{equation}\label{seconde partie de l estimation de mu 1}
\eta \sum_{k=1}^n
\big\| \mathcal P^{0 *,(k-1)\eta}
\big( \tilde L^{*} \Proba' - \tilde L^{*} \Proba' (1 |\mathbf e)  \big) \big\|_1
\; = \; 
\mathcal O \Big( \eta \sum_{k=1}^n
\ed^{-c (k-1) \eta} \epsilon 
\Big)
\; = \;
\mathcal O (\epsilon),
\end{equation}
where $c$ denotes some strictly positive constant. 
Putting \eqref{proof of mu 1 estimate obtained from the projection} and \eqref{seconde partie de l estimation de mu 1} in \eqref{expression for mu 1}, 
one ends up with 
\begin{equation}\label{expression finale de mu1}
\mu_1 
\; = \;  
\tau \sum_{k=1}^N
\gamma (e_k,e_{k+1})\multiplication 
\big(  \Proba' (\sigma_k \mathbf e) - \Proba' (\mathbf e) \big)
\; + \; \mu_1',
\end{equation}
with $\| \mu_1' \|_1 = \mathcal O (\tau^2)$, since $\eta < \epsilon < \tau^2$.

\paragraph{4. Bounding $\mu_2$.}
One here establishes that $\| \mu_2 \|_1  = \mathcal O (\tau^2)$.
Since $\|\mathcal P^{\epsilon*,s}\mu \|_1 \le \|\mu \|_1$ for every $s\ge 0$ and every measure $\mu$, one has
\begin{equation}\label{bounding mu2 decomp 1}
\| \mu_2 \|_1
\le
(\epsilon^{-1}\tau)
\eta \sum_{j=1}^n
\|
L^* {\mathcal P}^{0*,(j-1)\eta} \tilde L^* \Proba' 
\|_1 ,
\end{equation}
and, because of the particular form \eqref{def of P prime} of $\Proba'$, 
\begin{equation}\label{bounding mu2 decomp 1bis}
\|
L^*  {\mathcal P}^{0*,(j-1)\eta} \tilde L^* \Proba' 
\|_1
\;\le\; 
\sup_{u:\|u\|_\infty \le 1}\,
\max_{\substack{
\mathbf p \, : \, p_1^2 = 2\mathsf e_1 \\ 
\phantom{\mathbf p \, : \,} \dots \\ 
\phantom{\mathbf p \, : \,} p_N^2 = 2\mathsf e_{N}}} \,
\int_{\QQ} 
|\tilde L {\mathcal P}^{0,(j-1)\eta} L u  (\mathbf q,\mathbf p)| 
\, \dd \mathbf q.
\end{equation}
So let us take $u\in\Lp^\infty(\Phi)$ such that $\| u \|_\infty \le 1$, and $\mathbf p\in \PP^N$ such that $p_k^2 = 2\mathsf e_{k}$ for $1 \le k \le N$.
To lighten some further notations, let us also define the time
\begin{equation}\label{def of t}
t = (j-1)\eta.
\end{equation}
By Lemma \ref{lemma: L and tilde L}, 
\begin{equation*}
\tilde L {\mathcal P}^{0,t} L u 
\; = \;
\tilde L {\mathcal P}^{0,t} \tilde L u
\; + \;
\tilde L {\mathcal P}^{0,t} \chi_{\mathrm Z} u', 
\end{equation*}
with $\| u' \|_\infty = \mathcal O (1)$. 
To avoid to deal with too many constants, one will assume that actually $\| u' \|_\infty \le 1$.
Then, from the definition \eqref{def of L tilde} of $\tilde L$, 
\begin{align*}
|\tilde L {\mathcal P}^{0,t} \tilde L u |
\; \le & \;
\frac{1}{\eta^2} 
\Big|
\chi_{\mathrm{Coll}} \multiplication (\tilde{\mathcal P}^{\epsilon, \eta} - \tilde{\mathcal P}^{0,\eta}) 
{\mathcal P}^{0,t} \big(\chi_{\mathrm{Coll}} \multiplication (\tilde{\mathcal P}^{\epsilon, \eta} - \tilde{\mathcal P}^{0,\eta}) u\big)
\Big|\\
\le &\;
\frac{2}{\eta^2}
\Big(
\chi_{\mathrm{Coll}} \multiplication \tilde{\mathcal P}^{\epsilon,\eta} {\mathcal P}^{0,t} \chi_{\mathrm{Coll}} 
\; + \;  
\chi_{\mathrm{Coll}} \multiplication \tilde{\mathcal P}^{0,\eta} {\mathcal P}^{0,t} \chi_{\mathrm{Coll}} 
\Big)
\end{align*} 
and, similarly, 
\begin{equation*}
|\tilde L {\mathcal P}^{0,t} \chi_{\mathrm Z} u' |
\; \le \; 
\frac{2}{\eta}
\Big(
\chi_{\mathrm{Coll}} \multiplication \tilde{\mathcal P}^{\epsilon,\eta} {\mathcal P}^{0,t} \chi_{\mathrm{Z}} 
\; + \;  
\chi_{\mathrm{Coll}} \multiplication \tilde{\mathcal P}^{0,\eta} {\mathcal P}^{0,t} \chi_{\mathrm{Z}}
\Big).
\end{equation*}
Therefore
\begin{align}
\int_\QQ 
|\tilde L {\mathcal P}^{0,t} L u (\mathbf q, \mathbf p) | \, \dd \mathbf q
\; \le &
\;\;\;\;\;
\frac{2}{\eta^2}\int_\QQ \chi_{\mathrm{Coll}} (\mathbf q, \mathbf p) 
\multiplication\tilde{\mathcal P}^{\epsilon, \eta} {\mathcal P}^{0,t} \chi_{\mathrm{Coll}} (\mathbf q, \mathbf p) \, \dd \mathbf{q} \label{borne mu 2 terme I}\\
&+\; 
\frac{2}{\eta^2}\int_\QQ \chi_{\mathrm{Coll}} (\mathbf q, \mathbf p)
\multiplication \tilde{\mathcal P}^{0, \eta} {\mathcal P}^{0,t} \chi_{\mathrm{Coll}} (\mathbf q, \mathbf p)  \, \dd \mathbf{q} \label{borne mu 2 terme II}\\
&+\; 
\frac{2}{\eta}\int_\QQ \chi_{\mathrm{Coll}} (\mathbf q, \mathbf p) 
\multiplication \tilde{\mathcal P}^{\epsilon, \eta} {\mathcal P}^{0,t} \chi_{\mathrm Z}  (\mathbf q, \mathbf p)  \, \dd \mathbf{q} \label{borne mu 2 terme III}\\
&+\;
\frac{2}{\eta}\int_\QQ \chi_{\mathrm{Coll}} (\mathbf q, \mathbf p) 
\multiplication \tilde{\mathcal P}^{0, \eta} {\mathcal P}^{0,t} \chi_{\mathrm Z}  (\mathbf q, \mathbf p) \, \dd \mathbf{q}.\label{borne mu 2 terme IV}
\end{align} 

The four terms appearing in the right hand side of this inequality are bounded in a very similar way, and we will only deal in detail with the first one, 
indicating at the end the minor needed adaptations to bound the three others. 
We write
\begin{equation}
\int_\QQ \chi_{\mathrm{Coll}} 
\multiplication\tilde{\mathcal P}^{\epsilon, \eta} {\mathcal P}^{0,t} \chi_{\mathrm{Coll}}  \, \dd \mathbf{q}
\; \le \; 
\sum_{1 \le k,l \le N-1} 
\int_\QQ \chi_{\mathrm{Coll}_k} 
\multiplication\tilde{\mathcal P}^{\epsilon, \eta} {\mathcal P}^{0,t} \chi_{\mathrm{Coll}_l} \, \dd \mathbf{q}.
\end{equation}
It follows from (\ref{decomposition de T en particules independantes}-\ref{ecriture de l operateur Tk}) and then \eqref{second result uncoupled kernel} in the next section, that
\begin{align*}
{\mathcal P}^{0,t} \chi_{\mathrm{Coll}_l} (x_l,x_{l+1})
\; =& 
\;\;\;\;\;\,
\ed^{-2\lambda t} \multiplication  \chi_{\mathrm{Coll}_l} (\tilde X^0_l(x_l,t), \tilde X^0_{l+1}(x_{l+1},t)) \\
&
+ \; \ed^{-\lambda t} \int g(x_{l+1},x_{l+1}',t)\multiplication \chi_{\mathrm{Coll}_l} (\tilde X^0_l (x_l,t), x_{l+1}') \, \dd x_{l+1}'\\
&
+ \; \ed^{-\lambda t} \int g(x_l,x_l',t)\multiplication \chi_{\mathrm{Coll}_l}(x_l',\tilde X^0_{l+1} (x_{l+1},t)) \, \dd x_l'\\
&
+ \; \int g(x_l,x_l',t)\multiplication g(x_{l+1},x_{l+1}',t)\multiplication \chi_{\mathrm{Coll}_l}(x') \, \dd  x'_1 \, \dd x'_{l+1},
\end{align*}
where we have used the notation
\begin{equation*}
\int u \, \dd x_m \; = \; \sum_{p_m =\pm |p_m|} \int_{[0,1]} u \, \dd q_m, \qquad 1 \le m \le N.
\end{equation*}
The first term in the right hand side of this expression cannot be simplified further. 
As far as the second is concerned, 
by the bound $\mathrm{Coll}_l \subset \mathrm Z_l$ and the definition \eqref{def of Z k} of $\mathrm Z_l$, 
one obtains
\begin{equation*}
\chi_{\mathrm{Coll}_l} (\tilde X^0_l (x_l,t), x_{l+1}')
\; \le \; 
\chi_{[1 - \epsilon, 1]}\big( \tilde q^0_l(x_l,t) \big) \multiplication \chi_{[0,\mathsf p \eta]}\big(\big|q'_{l+1} - \tilde q^0_l(x_l,t) + 1 - \epsilon \big|\big).
\end{equation*}
Now, $g$ is uniformly bounded according to Lemma \ref{Lemma: uncoupled dynamics} in the next section, and so 
\begin{equation*}
\int g(x_{l+1},x_{l+1}',t)\multiplication \chi_{\mathrm{Coll}_l} (\tilde X^0_l (x_l,t), x_{l+1}') \, \dd x_{l+1}'
\; = \; 
\mathcal O (\eta) \multiplication \chi_{[1 - \epsilon, 1]}\big( \tilde q^0_l(x_l,t) \big) .
\end{equation*}
By a similar computation for the third term, and using \eqref{size of Zk} for the fourth one, one concludes that 
\begin{align}
 {\mathcal P}^{0,t} \chi_{\mathrm{Coll}_l} (x_l,x_{l+1})
\; \le \; 
&\ed^{-2\lambda t} \multiplication  \chi_{\mathrm{Coll}_l} (\tilde X^0_l(x_l,t), \tilde X^0_{l+1}(x_{l+1},t)) 
\nonumber \\
&+\; 
\mathcal O (\eta) \multiplication \chi_{[1 - \epsilon, 1]}\big( \tilde q^0_l(x_l,t) \big)
\; + \; 
\mathcal O (\eta) \multiplication \chi_{[0 , \epsilon ]}\big( \tilde q^0_{l+1}(x_{l+1},t) \big)
\; + \;  
\mathcal O(\eta \epsilon) \nonumber \\
\; := &\; 
\sum_{i=1}^4 \mathrm A_{i,j,l} (x_l , x_{l+1})
\label{bounding mu2 decomp4}
\end{align}
where the index $j$ comes from the fact that $t = (j-1)\eta$ by \eqref{def of t}. 

Defining then 
\begin{equation*}
\mathrm B_{i,j,k,l} (p_k,p_{k+1}) \; = \; 
\frac{1}{\eta^2} 
\int_\QQ \chi_{\mathrm{Coll}_k} \multiplication \mathcal P^{\epsilon, \eta}  \mathrm A_{i,j,l} \, \dd \mathbf q, 
\end{equation*}
and going backwards in the expressions (\ref{bounding mu2 decomp 1}-\ref{bounding mu2 decomp4}), 
one concludes that it will be enough to show that, 
for $i=1,2,3,4$, and for $1 \le k,l \le N-1$, one has 
\begin{equation}\label{proof to be shown}
\eta \sum_{j=1}^n \mathrm B_{i,j,k,l} (p_k,p_{k+1}) 
\; = \; 
\mathcal O(\tau \epsilon), 
\end{equation}
which in particular will be the case if one establishes that 
\begin{equation*}
\mathrm B_{i,j,k,l}  (p_k,p_{k+1})= \mathcal O (\epsilon^2).
\end{equation*}
Five cases will be analyzed separately: 
($i=1, l=k$), 
($i=1, l=k \pm 1$),
($i=1, | l-k | > 1$),
($i=2,3$), 
($i=4$). 
The diophantine condition on the velocities only needs to be used in the case ($i=1, l=k$). 

\vspace{0.2cm}
\noindent
\emph{1. Bounding $\mathrm B_{1,j,k,k}$.}
One has
\begin{equation}\label{starting bound on beta 1jjk}
\mathrm B_{1,j,k,k} 
\;\le\; 
\frac{\ed^{-2\lambda t}}{\eta^2} \int_\QQ
\chi_{\mathrm{Coll}_k}(x)
\multiplication
\chi_{\mathrm{Coll}_k} 
\big( 
\tilde X^0_k(\tilde X^\epsilon_k(x,\eta),t), \tilde X^0_{k+1}(\tilde X^\epsilon_{k+1}(x,\eta),t)
\big) 
\, \dd \mathbf q.
\end{equation}
Let us take some $x\in \mathrm{Coll}_k$.
The first point to observe is that, 
for the deterministic realization $\tilde X^\epsilon$ of the process $X^\epsilon$,
collisions involving a given particle are spaced by time intervals of order 1 at least.  
There exists therefore a constant $c>0$ such that, for $t=(j-1)\eta \le c$, 
\begin{equation*}
\big( 
\tilde X^0_k(\tilde X^\epsilon_k(x,\eta),t), \tilde X^0_{k+1}(\tilde X^\epsilon_{k+1}(x,\eta),t)
\big) 
\; \notin \; 
\mathrm{Coll}_k, 
\end{equation*}
so that  $\mathrm B_{1,j,k,k} = 0$ in that case.
Let then suppose $t\ge c$, and let us just bound the second factor in the integrand of \eqref{starting bound on beta 1jjk} by
\begin{equation*}
\chi_{\mathrm{Coll}_k} (x_k,x_{k+1})
\;\le\;
\chi_{[1 - 2\epsilon,1] \times [0, 2\epsilon]} (q_k,q_{k+1}),
\end{equation*}
valid since $\mathrm{Coll}_k \subset \mathrm Z_k$, and by the definition \eqref{def of Z k} of $\mathrm Z_k$.
Let then $t^* > c$ be the smallest time such that 
\begin{equation*}
\big( \tilde q_k^0 (\tilde X^\epsilon_k(x,\eta),t^*), \tilde q_{k+1}^0 (\tilde X^\epsilon_{k+1}(x,\eta),t^*) \big)
\in 
[1 - 2\epsilon,1] \times [0,2\epsilon].
\end{equation*}
There has to be two integers $r,r'\ne 0$ such that
\begin{equation*}
t^* 
= 
\frac{2r}{p_k} + \mathcal O (\epsilon)
=
\frac{2r'}{p_{k+1}} + \mathcal O (\epsilon),
\end{equation*}
and so
\begin{equation*}
\Big| \frac{p_k}{p_{k+1}} - \frac{r}{r'} \Big|
\; = \; \mathcal O (\epsilon)
\quad \text{and} \quad
\Big| \frac{p_{k+1}}{p_{k}} - \frac{r'}{r} \Big|
\; = \;
\mathcal O(\epsilon).
\end{equation*}
As said after \eqref{bounding mu2 decomp 1bis}, $p_k^2 = 2 \mathsf e_k$ and $p_{k+1}^2  = 2 \mathsf e_{k+1}$, and, by hypothesis, 
the quotients $p_k/p_{k+1}$ and $p_{k+1}/p_k$ are thus diophantine by hypothesis. 
This condition guaranties the existence of a constant $\mathrm c >0$ such that $r,r' \,\ge\, c \, \epsilon^{-1/\beta}$, 
and so one has also $t^* > c' \, \epsilon^{-1/\beta}$ for some $c' > 0$.
Since $\mathrm B_{1,j,k,k} = 0$ as long as $t = (j-1)\eta \le t^*$, and since $\eta> \epsilon^4$, one concludes that  
\begin{equation}\label{proof case i 1 first}
\mathrm B_{1,j,k,k} 
\; \le \; 
\frac{  \ed^{-2\lambda \, c' \, \epsilon^{-1/\beta}}  }{\eta^2}
\; = \; \mathcal O (\epsilon^2).
\end{equation}

\vspace{0.2cm}
\noindent
\emph{2. Bounding $\eta \sum_{j=1}^{n}\mathrm B_{1,j,k,k\pm 1}$.}
The two cases, $l=k+1$ and $l=k-1$, are analogous, and one will treat the first one only.
One has
\begin{equation*}
\mathrm B_{1,j,k,k+1} 
\;\le\; 
\frac{\mathrm C }{\eta^2} \int_\QQ
\chi_{\mathrm{Coll}_k}(x)
\multiplication
\chi_{\mathrm{Coll}_{k+1}} 
\big( 
\tilde X^0_{k+1}(\tilde X^\epsilon_{k+1}(x,\eta),t), \tilde X^0_{k+2}(\tilde X^\epsilon_{k+2}(x,\eta),t)
\big) 
\, \dd \mathbf q.
\end{equation*}
Integrating over $q_k$ and $q_{k+1}$, one gets
\begin{equation*}
\mathrm B_{1,j,k,k+1} 
\;\le\; 
\frac{\mathrm C \,\epsilon}{\eta} 
\sup_{\substack{q_k\in [1 -2 \epsilon,1] \\ q_{k+1}\in [0,2\epsilon]}} \;
\int
\chi_{\mathrm{Z}_{k+1}} 
\big( 
\tilde X^0_{k+1}(\tilde X^\epsilon_{k+1}(x,\eta),t), \tilde X^0_{k+2}(\tilde X^\epsilon_{k+2}(x,\eta),t)
\big) 
\, \dd q_{k+2} \dd q_{k+3} 
\end{equation*} 
One may obviously insert the factor $\chi_{\mathrm{Z}_{k+2}} + (1 - \chi_{\mathrm{Z}_{k+2}})$ in this integral, and split it accordingly. 
In the term containing the factor $\chi_{\mathrm{Z}_{k+2}}$, one may just use the bound $\chi_{\mathrm{Z}_{k+1}} (\dots ) \le 1$ so that this term is $\mathcal O (\eta \epsilon)$, 
whereas, in the term containing the factor $(1 - \chi_{\mathrm{Z}_{k+2}})$, one may replace $\tilde{X}_{k+2}^\epsilon$ by $\tilde{X}^0_{k+2}$. At the end, 
\begin{equation*}
\mathrm B_{1,j,k,k+1} 
\;\le\; 
\mathrm C \epsilon^2 \; + \; \frac{\mathrm C \epsilon}{\eta} \sup_{\substack{q_k\in [1 -2 \epsilon,1] \\ q_{k+1}\in [0,2\epsilon]}} \; 
\int
\chi_{\mathrm{Z}_{k+1}}
\big( 
\tilde X^0_{k+1}(\tilde X^\epsilon_{k+1}(x,\eta),t), \tilde X^0_{k+2}(x,t+\eta)
\big) 
\, \dd q_{k+2} .
\end{equation*}
For the integrand to not be 0, the two following conditions need to be fulfilled: 
\begin{align*}
& \tilde q^0_{k+1} (\tilde X^\epsilon_{k+1}(x_k,x_{k+1},\eta),t) 
\; \in \;  [1 - 2 \epsilon, 1], \\
& \big| \tilde q^0_{k+2}(x_{k+2},t+\eta) - \tilde X^0_{k+1}(\tilde q^\epsilon_{k+1}(x_k,x_{k+1},\eta),t) \big|
\; = \; \mathcal O(\eta). 
\end{align*}
The uncoupled dynamics is such that, for any $a\in [0,1]$, any $s\ge 0$, 
\begin{equation*}
\mathrm{Leb} \big(  | \tilde q^0_{k+2}(x_{k+2},s) -a | = \mathcal O (\eta) \big) 
\; = \; \mathcal O (\eta).
\end{equation*}
So, integrating over $q_{k+2}$, 
\begin{equation*}
\mathrm B_{1,j,k,k+1} 
\;\le\; 
\mathrm C \epsilon^2 \; + \; \mathrm C \epsilon \sup_{\substack{q_k\in [1 -2 \epsilon,1] \\ q_{k+1}\in [0,2\epsilon]}} \; 
\chi_{[1-2\epsilon,1]} \big( \tilde q^0_{k+1} (\tilde X^\epsilon_{k+1}(x_k,x_{k+1},\eta),(\eta-1)j)  \big),
\end{equation*}
where one has used the definition \eqref{def of t} of $t$. 
Let us thus fix some $q_k\in [1 -2 \epsilon,1]$ and some $q_{k+1}\in [0,2\epsilon]$.
Summing $\mathrm B_{1,j,k,k+1} - \mathrm C\epsilon^2$ over $j$, one sees that only a proportion $\epsilon$ of the terms are non zero, and so 
\begin{equation}
\eta \sum_{j=1}^n \mathrm B_{1,j,k,k+1}
\; = \;
\mathcal O ( \epsilon \tau ).
\end{equation}

\vspace{0.2cm}
\noindent
\emph{3. Bounding $\mathrm B_{1,j,k,l}$, $|l-k| \ge 1$.}
One has
\begin{equation*}
\mathrm B_{1,j,k,l} 
\;\le\; 
\frac{1}{\eta^2} \int_\QQ
\chi_{\mathrm{Coll}_k}(x)
\multiplication
\chi_{\mathrm{Coll}_{l}} 
\big( 
\tilde X^0_{l}(\tilde X^\epsilon_{l}(x,\eta),t), \tilde X^0_{l+1}(\tilde X^\epsilon_{l+1}(x,\eta),t)
\big) 
\, \dd \mathbf q.
\end{equation*}
Integrating over the variables $q_k,q_{k+1}$, one gets, 
\begin{equation*}
\mathrm B_{1,j,k,l} 
\;\le\; 
\frac{\mathrm C\, \epsilon}{\eta} 
\sup_{\substack{q_k\in [1 -2 \epsilon,1] \\ q_{k+1}\in [0,2\epsilon]}} \;
\int
\chi_{\mathrm{Coll}_{l}} 
\big( 
\tilde X^0_{l}(\tilde X^\epsilon_{l}(x,\eta),t), \tilde X^0_{l+1}(\tilde X^\epsilon_{l+1}(x,\eta),t)
\big) 
\, \dd \mathbf q_{k,k+1}
\end{equation*}
where $\dd \mathbf q_{k,k+1} = \dd q_1 \dots \dd q_{k-1} \dd q_{k+2} \dots \dd q_{N}$.
One may obviously insert the factor
\begin{equation*}
\chi_{\mathrm{Z}_{l-1} \cup  \mathrm{Z}_{l} \cup  \mathrm{Z}_{l+1} } (x)\; + \; 
\big(1 - \chi_{\mathrm{Z}_{l-1} \cup  \mathrm{Z}_{l} \cup  \mathrm{Z}_{l+1} } (x)\big)
\end{equation*} 
into this last integral, and split it accordingly. 
In the term involving the factor $\chi_{\mathrm{Z}_{l-1} \cup  \mathrm{Z}_{l} \cup  \mathrm{Z}_{l+1} }$, one just uses the bound 
$\chi_{\mathrm{Coll}_l} (\dots) \le 1$ to obtain that this term is $\mathcal O (\eta \epsilon)$, 
whereas in the other one, one may replace $\tilde X^\epsilon$ by $\tilde X^0$. At the end, 
\begin{equation*}
\mathrm B_{1,j,k,l} 
\;\le\; 
\mathrm C \, \epsilon^2 
\; + \;
\frac{\mathrm C \, \epsilon}{\eta} \int    
\chi_{\mathrm{Coll}_{k+1}} 
\big( 
\tilde X^0_{l}(x,t+\eta), \tilde X^0_{l+1}(x,t+\eta)
\big)
\, \dd \mathbf q_{k,k+1}, 
\end{equation*}
and this last integral is $\mathcal O (\eta \epsilon)$. So 
\begin{equation}\label{borne sur B 1 j k l}
\mathrm B_{1,j,k,l} 
\;=\; 
\mathcal O (\epsilon^2).
\end{equation}

\vspace{0.2cm}
\noindent
\emph{4. Bounding $\eta \sum_{j=1}^n\mathrm B_{2,j,k,l}$ and $\eta \sum_{j=1}^n\mathrm B_{3,j,k,l}$.}
These two cases are similar, and we treat only the first one:
\begin{align*}
\eta \sum_{j=1}^n\mathrm B_{2,j,k,l}
\;\le &\;
\frac{\mathrm C}{\eta} \int_\QQ \chi_{\mathrm{Coll}_k}(x)
\multiplication \tilde{\mathsf P}^{\epsilon, \eta} \eta\sum_{j=1}^n  \chi_{[1-\epsilon, 1]}\big(\tilde X_l(x_l,(j-1)\eta)\big) \, \dd \mathbf q \nonumber\\
\le &\;
\mathrm C \, \epsilon
\multiplication
\sup_{x\in\Phi}  \eta\sum_{j=1}^n  \chi_{[1-\epsilon, 1]}\big(\tilde X_l(x_l,(j-1)\eta)\big).
\end{align*}
Since, for any $x\in\Phi$, the number of non zero terms in the sum is $\mathcal O (n\epsilon)$, 
\begin{equation}\label{la seconde borne pour la preuve de mu2}
\eta \sum_{j=1}^n\mathrm B_{2,j,k,l}
\; = \; 
\mathcal O (\epsilon \tau).
\end{equation}

\vspace{0.2cm}
\noindent
\emph{5. Bounding $\mathrm B_{4,j,k,l}$.}
\begin{equation}\label{5e element ds la preuve de mu 2}
\mathrm B_{4,j,k,l}
\;\le\;
\frac{\mathrm C}{\eta^2} \int_Q \chi_{\mathrm{Coll}_k}(x) \multiplication \tilde{\mathcal{P}}^{\epsilon,\eta} \epsilon \eta \, \dd \mathbf q 
\;=\;
\mathcal O ( \epsilon^2)
\end{equation}

\vspace{0.2cm}
\noindent
\emph{6. Concluding the bound on $\mu_2$.}
The bounds (\ref{proof case i 1 first}-\ref{5e element ds la preuve de mu 2})
show that \eqref{proof to be shown} holds for $i=1,2,3,4 $ and $1 \le k,l \le N-1$.
However, as said after (\ref{borne mu 2 terme I}-\ref{borne mu 2 terme IV}), one still needs to explain
how to adapt the proof when one starts with one of the terms (\ref{borne mu 2 terme II}-\ref{borne mu 2 terme IV}), 
instead of \eqref{borne mu 2 terme I}.
If one starts from \eqref{borne mu 2 terme II}, all the previous arguments actually still hold.
If instead one starts from \eqref{borne mu 2 terme III}, 
the only place where some non-immediate adaptation is required,
concerns the bound on $\mathrm B_{1,j,k,k}$ for $t =(j-1)\eta < c$, where $c$ is the constant introduced just after \eqref{starting bound on beta 1jjk}. 
Indeed, for $x\in \mathrm{Coll}$, it can be that 
\begin{equation*}
\big( 
\tilde X^0_k(\tilde X^\epsilon_k(x,\eta),t), \tilde X^0_{k+1}(\tilde X^\epsilon_{k+1}(x,\eta),t)
\big) 
\; \in \; 
\mathrm Z_k
\end{equation*}
for a finite and constant number of times $t=(j-1)\eta < c$. 
Since now the factor $1/\eta^2$ in the beginning of right hand side of \eqref{starting bound on beta 1jjk} may be replaced by a factor $1 / \eta$, 
one have, for such times, only the bound  
\begin{equation*}
\mathrm B_{1,j,k,k}
\; \le \;
\frac{1}{\eta} \int_\QQ \chi_{\mathrm{Coll}_k} (x) \, \dd \mathbf q.
\; = \;
\mathcal O (\epsilon)
\end{equation*} 
But then, inserting this in  \eqref{proof to be shown}, one gets $\eta \sum_{j} \mathrm B_{1,j,k,k}  = \mathcal O(\eta\epsilon)=\mathcal O (\epsilon \tau)$, 
since the sum contains only a finite and constant number of non-zero terms.
The case of \eqref{borne mu 2 terme IV} is analogous. $\square$

\paragraph{5. The case $N' < N$.}
If $N' < N$, the arguments leading from \eqref{starting bound on beta 1jjk} to \eqref{proof case i 1 first} do not hold anymore, and this is the only place where a problem occurs. 
Indeed one assumed there that all the energies were different in order to make use of the diophantine condition. 
This problem is clearly superficial, since we do not care about collisions between particles having the same energy,  
and it can indeed be fixed by a slight modification of the proof.

Let us go back to the general strategy explained in Subsection \ref{subsection: method}, 
and let us again start from a measure $\mathsf P$ assigning to each particle a given energy.
When Duhamel's formula is used for the first time, in \eqref{Duhamel 1er ordre}, instead of taking $S$ to be the operator $\mathcal P^{0*,\eta}$ decoupling all the particles, 
one can take an operator which decouples only blocs of particles having a given energy. 
Up to an error of order $\epsilon^2$, this measure is still invariant under this dynamics, and this is in fact all what we need to proceed.
When Duhamel's formula is used for the second time, in \eqref{method 2}, one still takes $S$ to be $\mathcal P^{0*,\eta}$ however, 
since otherwise one could not make use of the properties obtained in the Section \ref{section: uncoupled dynamics}. 

It can then be checked that the proof essentially remains unchanged but that, when bounding $\mu_2$, 
one may now assume that recollisions only occur between particles of different energies, allowing thus the use of the diophanitine condition. 
We decided however to not implement these changes explicitly to avoid even more heavy notations.

\section{Uncoupled dynamics}\label{section: uncoupled dynamics}

We analyze here the dynamics when $\epsilon = 0$, so that all the particles evolve independently.

\subsection{$N = 1$ particle}

Take $N = 1$, and let $\mathsf e_1$ be the energy of the only particle.
Let $\overline X = (\overline q, \overline p)$ be the chain on $\R \times \PP = \R \times \{-\mathsf e_1,\mathsf e_1 \}$ 
solving the equations
\begin{equation*}
\dd \overline q \; = \; \overline p \, \dd t, 
\qquad
\dd \overline p \; = \; - 2 \overline q \, \dd \mathrm N,
\end{equation*}
where $\mathrm N$ is a Poisson process of parameter $\lambda >0$.
For a given smooth function $v$ on $\R \times \PP$, let 
\begin{equation*}
u (x,t) = \mathsf E \big(v\circ \overline X(x,t) \big),
\end{equation*}
which, for $t\ge 0$, satisfies the equation
\begin{equation*}
\partial_t u (q,p,t) = p\, \partial_q  u (q,p,t) + \lambda \big(  u(q,-p,t) -  u(q,p,t)\big).
\end{equation*} 
It is lengthy but straightforward to check that $u$ may be written as
\begin{equation*}
u(q,p,t)
=
\sum_{p' = \pm p} \int_{-\infty}^{+\infty} \rho (q,p,q',p',t)\, v(q',p')\, \dd q'
\end{equation*} 
where the probability measure $\rho$ is given by 
\begin{equation}\label{uncoupled kernel on R}
\rho(q,p,q',p',t)
=
\ed^{-\lambda t} \sum_{n=0}^\infty \frac{(\lambda t)^n}{n !} \rho_n (q,p,q',p',t)
\end{equation}
with
\begin{eqnarray*}
\rho_0(q,p,q',p',t) 
&= &
\chi_{\{0 \}}(p - p')
\delta_0(pt +q - q'), \\
\rho_n (q,p,q',p',t)
&=&
\chi_{[-|p|t,|p|t]} (q'-q)
\chi_{\{0 \}}(p - p')
\frac{n! \,\Big( 1+ \frac{q'-q}{p t} \Big)^{\frac{n}{2}}\Big(  1- \frac{q'-q}{p t} \Big)^{\frac{n}{2}-1}}{|p|t \, 2^n \Big(\frac{n}{2}\Big)! \Big( \frac{n}{2}-1\Big)!}, 
\quad n \ge 2 \text{ even}, \\
\rho_n (q,p,q',p',t)
&=&
\chi_{[-|p|t,|p|t]} (q'-q)
\chi_{\{0 \}}(p + p')
\frac{n! \,\Big( 1+ \frac{q'-q}{p t} \Big)^{\frac{n-1}{2}}\Big(  1- \frac{q'-q}{p t} \Big)^{\frac{n-1}{2}}}{|p|t \, 2^n\Big(\frac{n-1}{2}\Big)! \Big( \frac{n-1}{2}\Big)!},
\quad n \ge 1 \text{ uneven}.
\end{eqnarray*}

This allows us to give an expression for the kernel of the evolution operator $\mathcal P^{0,t}$ 
of the chain $X$ on $\Phi = [0,1]\times \{-\mathsf e_1, \mathsf e_1 \}$ defined by (\ref{equation of the motion 1}-\ref{equation of the motion 2}).
Given a function $u$ on $\Phi$, and given $t\ge 0$, let us write 
\begin{equation*}
\mathcal P^{0,t} u (x)
=
\sum_{p' = \pm p} \int_0^1 f(q,p,q',p',t) \, u(q',p') \, \dd q'.
\end{equation*}
The trajectories of $X$ are obtained from these of $\overline X$ by reflecting them against the boundaries at $q=0$ and $q=1$, and therefore
\begin{equation}\label{uncoupled kernel}
f(q,p,q',p',t)
=
\sum_{k\in \Z, \text{even}}
\rho (q, p , q' + k, p', t)
+\sum_{k\in \Z, \text{uneven}}
\rho (q, p , 1 - q' + k, -p', t).
\end{equation}

\begin{Lemma}\label{Lemma: uncoupled dynamics}
One has this.

\noindent
1.
(Doeblin's condition) There exists $\alpha >0$ and $t_0 \ge 0$ such that 
$f(x,x',t_0) \ge \frac{\alpha}{2}$
for every $x,x'\in \Phi$.

\noindent 
2.
There exists a function $g\in \Lp^\infty (\Phi^2 \times \R_+)$ such that 
\begin{equation}\label{second result uncoupled kernel}
f(x,x',t)
=
\ed^{-\lambda t}
\chi_{\{ \tilde{p} (x,t) \}} (p') \, \delta_{\tilde{q} (x,t)} (q')
+ 
g(x,x',t).
\end{equation}
\end{Lemma}

\Proof
The first part is directly established using (\ref{uncoupled kernel on R}) and (\ref{uncoupled kernel}).
Let us show the second one. 
By (\ref{uncoupled kernel}), the first term in the right hand side of (\ref{second result uncoupled kernel}) directly comes from $\rho_0$ in (\ref{uncoupled kernel on R}).
Then, inserting (\ref{uncoupled kernel on R}) in (\ref{uncoupled kernel}), and putting the sums over $k$ inside the sums over $n$, 
one sees that it suffices to find a constant $\mathrm C < + \infty$ such that, 
for every $p\in \PP$, for every $z\in\R$ and for every $n\ge 1$, one has
\begin{equation}\label{preuve lemme dynamique non couplee}
 \frac{n!}{2^n \Big(\frac{n}{2}\Big)! \Big( \frac{n}{2}-1\Big)!} \,
\frac{1}{|p| t} 
\hspace{-0.3cm}
\sum_{\stackrel{k\in \Z}{|z+k| \le |p|t}}
\Big( 1+ \frac{z+k}{p t} \Big)^{\frac{n}{2}}\Big(  1- \frac{z+k}{p t} \Big)^{\frac{n}{2}-1}
\le
\mathrm C,
\end{equation}
if $n$ is even, and
 \begin{equation}\label{preuve lemme dynamique non couplee bis}
\frac{n!}{2^n \Big(\frac{n-1}{2}\Big)! \Big( \frac{n-1}{2}\Big)!} \,
\frac{1}{|p| t}
\hspace{-0.3cm}
\sum_{\stackrel{k\in \Z}{|z+k|\le |p|t}}
\Big( 1+ \frac{z+k}{p t} \Big)^{\frac{n-1}{2}}\Big(  1- \frac{z+k}{p t} \Big)^{\frac{n-1}{2}}
\le
\mathrm C,
\end{equation}
if $n$ is uneven.
Let us show \eqref{preuve lemme dynamique non couplee} ; the proof of \eqref{preuve lemme dynamique non couplee bis} is analogous.
One finds $c >0$ such that, for every $x\in [-1,1]$,  
\begin{equation*}
(1+x)^{n/2}(1-x)^{n/2 - 1} 
\; \le \;
2 (1 - x^2)^{n/2 - 1}
\; = \; 
\mathcal O \big( \ed^{-c n x^2 } \big).
\end{equation*}
But 
\begin{equation*}
\frac{1}{|p| t} 
\hspace{-0.3cm}
\sum_{\stackrel{k\in \Z}{|z+k| \le |p|t}}
\ed^{-cn \big(\frac{z+k}{p t}\big)^2} 
\; = \;  
\mathcal O (n^{-1/2}).
\end{equation*}
On the other hand, one establishes by means of Stirling's formula that the combinatoric factor in the left hand side of \eqref{preuve lemme dynamique non couplee}
is $\mathcal O (n^{1/2})$. This completes the proof. $\square$

\subsection{$N \ge 1$ particles}
Lemma \ref{Lemma: uncoupled dynamics} has direct implications to the case $N \ge N' \ge 1$. 
Since the particles evolve independently, the operator $\mathcal P^{0,t}$ can be written as the tensor product
\begin{equation}\label{decomposition de T en particules independantes}
\mathcal P^{0,t}
=
\mathcal P_1^{t} \dots \mathcal P_N^t,
\end{equation} 
where, for $1\le k \le N$, $\mathcal P_k^t$ is the evolution operator of a chain which moves only particle $k$ and let fixed the other ones: 
\begin{equation}\label{ecriture de l operateur Tk}
\mathcal P_k^t u (x)
=
\sum_{p_k' = \pm p_k}
\int_{0}^{1}
f (q_k ,p_k, q_k',p_k',t)\,
u(q_1, \dots, q_k', \dots , q_N, p_1, \dots , p_{k}',\dots , p_N) \, \dd q_k', 
\end{equation}
with $f$ defined by \eqref{uncoupled kernel}.
Now, Doeblin's condition (the first part of Lemma \ref{Lemma: uncoupled dynamics}) means that 
the operators $\mathcal P_k^{0*,t_0}$ have a spectral gap for the $\|\multiplication\|_1$-norm. 
By the tensor product decomposition \eqref{decomposition de T en particules independantes}, 
one concludes thus that there exist constants $\mathrm C < + \infty$ and $c>0$ such that, for every measure $\mu$ on $\Phi$, 
\begin{equation}\label{uncoupled dynamics result 1}
\big\| \mathcal P^{0*,t}\mu -  \mu (1 | \mathbf e)   \big\|_1
\; \le \;  
\mathrm C \, \ed^{-c t} \| \mu \|_1,
\end{equation}
with $\mu(1 | \mathbf e)$ defined by \eqref{def projection sur S 0},
$\mathbf e = (e_1, \dots , e_N)$, 
and $e_k$ the energy of particle $k$.

\pagebreak
\section{Figures}\label{section: figures}

\noindent\textbf{Acknowledgments.}
I thank S. Olla for his interest in this work as well as for having introduced me to the subject. 
I thank C. Liverani for useful discussions. 
I thank the European Advanced Grant Macroscopic Laws and Dynamical Systems (MALADY) (ERC AdG 246953) for financial support.

\pagebreak

\end{document}